\pdfoutput=1  
\documentclass[12pt]{article}
\usepackage{amsmath,graphicx}
\usepackage{hyperref}
\usepackage{caption}
\usepackage{natbib}
\usepackage{multirow}
\usepackage{booktabs}
\usepackage{graphicx}
\usepackage{lmodern}			
\usepackage[T1]{fontenc}		
\usepackage[utf8]{inputenc}		
\usepackage{indentfirst}		
\usepackage{nomencl} 			
\usepackage{color}				
\usepackage[dvipsnames]{xcolor}
\usepackage{microtype} 			

\title{Lithospheric Structure of Venusian Crustal Plateaus}

\author{J. S. Maia$^1$, M. A. Wieczorek$^1$}

\date{\small $^1$Universit{\'e} C{\^o}te d'Azur, Observatoire de la C{\^o}te d'Azur, CNRS, Laboratoire Lagrange, France}

\begin{document}
\maketitle


\begin{abstract}

Crustal plateaus are Venusian highlands characterized by tectonized terrains. It is commonly interpreted that their topography is isostatically supported and that they represent fossils of an extinct tectonic regime. Using gravity and topography we perform a comprehensive investigation of the lithospheric structure of six crustal plateaus. We computed the admittance (gravity to topography wavelength-dependent ratio)  for each region and compared them to modeled admittances. Three compensation scenarios were tested: Airy isostasy, a surface-loading flexural model, and a flexural model with surface and subsurface loads. Our results show that the topography of most plateaus is supported by crustal thickening and that the addition of a mantle support component is not necessary at the investigated wavelengths. The elastic thickness was constrained to be less than 35 km with a best-fitting average of 15 km, confirming that these regions are consistent with an isostatic regime. The average crustal thickness of the plateaus ranges from 15 to 34 km, and if they are in Airy isostasy, this implies that the global average crustal thickness of Venus is about 20 km. Phoebe Regio is the sole exception of our analysis in that crustal thicknesses that are compatible with the other plateaus are obtained only when a buoyant layer is included. Heat flow estimations computed from the elastic thickness indicate that the plateaus formed under higher heat flow conditions compared to the current global average and could have caused localized melting. Present-day heat flow predictions suggest that eclogitization could occur where the crust is thickest.

\vspace{4pt}

\begin{center}
  \textbf{Plain Language Summary}
\end{center}
Crustal plateaus are large and highly deformed highlands observed uniquely on Venus. It is generally assumed that these features are in an Airy isostasy regime, where the weight of the topography is balanced by a crustal root floating over a higher-density mantle, and that the geologic processes responsible for their formation are no longer taking place.  However, their origin, structure, and evolution are still topics of great debate. This study aims to investigate the lithosphere and crustal structure of six of these regions making use gravity and topography data. We use the observed topography and geophysical models of the lithosphere to predict the gravitational signature of each region which is then compared to the observed gravity. With this analysis we constrained the thickness of the elastic lithosphere, which is a proxy for the rigidity of the lithosphere, and estimated the crustal thickness of these regions. We found that the elastic thickness of the crustal plateaus is in general very small, which suggests that the common assumption of Airy isostasy is overall valid. Moreover, we found that the crustal thickness of these regions is about 25 km on average.

\end{abstract}

%
%

\section{Introduction}

Crustal plateaus, also called plateau highlands, are prominent geologic features on Venus, with roughly circular planforms and diameters ranging from 1500 to 2500 km. They present a steep-sided topography reaching 2 to 4 km of altitude above the surrounding plains, with the highest elevations generally closer to the margins. The surface of the plateaus is dominated by tessera terrains, that are characterized by complex tectonic fabrics which indicate multiple stages of deformation recording both extensional and contractional events \cite[e.g.][]{Bindschadler1992, Hansen1996}. These terrains, which can also be found as low-lying patches within the plains called inliers, cover roughly 8\% of the surface of Venus and are stratigraphically the oldest surfaces on the planet \citep{Ivanov1996}. Given their relative older age, crustal plateaus have recorded a significant fraction of Venus' geologic history. Thus, investigating their structure and formation mechanism is crucial to decipher the early tectonic and geodynamic processes of the planet.

The origin of crustal plateaus is still a matter of debate. It is well established that the high topography observed is associated with thickening of the crust, but which processes caused this thickening are not quite understood. The high-resolution full-coverage radar imagery from the Magellan mission resulted in the first detailed geological maps of these regions. Different interpretations of their tectonic evolution developed into the ``hotspot-coldspot controversy'' \citep{PhillipsHansen1994}, a debate that carried throughout the 1990s. A first class of model considered upwelling and assumed that the crust was initially uplifted by a mantle plume and thickened by magmatic accretion due to plume-related partial melting, followed by cooling and subsidence \cite[e.g.][]{Phillips1998, GHENT1999, Hansen2000}. The opposing scenario proposed that crustal plateaus are regions of downwelling related to coldspots in the mantle \cite[e.g.][]{Bindschadler1991, Bindschadler1992b, GILMORE2000}. In this case, the crustal thickening is caused by horizontal shortening related to compressional stresses. However, \cite{kidder1996} have shown that the formation of plateaus through subsolidus crustal thickening over downwellings would require an excessively long time scale (from 1 to 4 billion years). On the other hand, the upwelling model has difficulties to accommodate the pervasive contractional tectonics observed at these regions.

An alternative interpretation related to large asteroid impacts was later proposed by \cite{Hansen2006} for the origin of the crustal plateaus. In this scenario, the collision of large bolides onto the surface of Venus are able to partially melt the crust and upper mantle generating huge lava ponds. The high relief in this model is supported by a depleted upper mantle residuum which is more buoyant and stronger than adjacent undepleted mantle. This model, however, has difficulties to explain the significant amount of shortening near the plateau margins \citep{Romeo2008}. Later, \cite{Romeo2008} and \cite{Romeo2011} suggested an alternative scenario where the crustal plateaus formed under similar conditions as the continental crust on Earth. According to this hypothesis, plateaus and tessera inliers represent buoyant areas of felsic composition with respect to the surroundings, which survived a putative global lithospheric foundering event around 500 Ma ago \cite[e.g.,][]{Turcotte1993, Turcotte1999, Weller2020}. The intense tectonic activity related to this event would contribute to building up the plateaus by compression. Consistent with this hypothesis, thermal emissivity data indicate that crustal plateaus could be associated with a more felsic composition, analogous to the composition of continental crust \citep{Hashimoto2008, Gilmore2015}.

The earliests investigations using gravity data on Venus showed that crustal plateaus are associated with small positive gravity anomalies and low gravity to topography ratios. The volcanic rises, such as Atla and Beta regiones, in contrast are associated with high gravity anomalies and high gravity-topography ratios \citep{smrekar1991, Bindschadler1992b, grimm_1994, kuncinskas_1994, moore_1997, Simons1994, simons_1997}. These studies led to the widely accepted interpretation that the topography of the crustal plateaus is isostatically compensated by thick crustal roots. Consequently, most investigations of gravity and topography in the Magellan era considered an Airy isostasy regime in order to estimate the crustal thickness of the plateaus. For this purpose, \cite{smrekar1991}, \cite{kuncinskas_1994}, and  \cite{moore_1997} adopted spatial analysis techniques (geoid-to-topography ratios). Meanwhile, \cite{grimm_1994} and \cite{Simons1994, simons_1997} investigated localized spectral admittances, which are wavelength-dependent gravity-topography ratios, making use of early developed spatio-spectral localization techniques. It is important to emphasize that none of these works made use of the most recent gravity model from \cite{Konopliv1999} which is the highest-resolution model publicly available today \cite[see][for a review]{Sjogren1997}. Moreover, many of these studies did not try to quantify the flexural strength of the lithosphere, and instead simply assumed an isostatic regime, where the lithosphere has zero strength. \cite{simons_1997} performed preliminary tests on the elastic support of the topography of Venusian features making use of a top-loading flexural model and concluded that the crustal plateaus are generally consistent with elastic thicknesses lower than 20 km.

The few recent gravity studies on Venus have mostly focussed on constructing global crustal thickness and elastic thickness maps. \cite{anderson_2006} systematically computed localized spectral admittances across the planet using the wavelet technique introduced by \cite{simons_1997} and divided them into spectral classes. Crustal thickness and elastic thicknesses were estimated for each spectral class considering top, bottom and ``hot-spot'' loading models. \cite{james_2013} investigated the crustal thickness of Venus using spatial domain geoid-to-topography ratios and created a global model considering  crustal thickness variations and dynamic compensation at depth arising from the mantle. Global maps of elastic thickness variations for the Moon, Mars and Venus were presented by \cite{Audet2014} using a spherical wavelet analysis and a flexural loading model of the lithosphere. A crustal thickness map based on the modeling technique introduced by \cite{Wieczorek1998} was constructed by \cite{jimenezdiaz_2015}. These authors then used wavelet transforms to perform localized spectral analyses in order to provide elastic thickness estimations considering a lithospheric model with surface and subsurface loads and using the crustal thickness values previously estimated. In addition, several gravity studies have focused on coronae, which are circular volcano-tectonic structures uniquely present on Venus, adopting top and bottom loading flexural models to investigate their crustal and elastic thicknesses \citep{smrekar2003, Hoogenboom2004, Hoogenboom2005}.

Since the time of the pioneering studies of the Magellan era, several advances have been made in the techniques that are used to analyze the gravity signal arising from the lithosphere. These include improved spatio-spectral localization techniques, notably the spherical multitaper spectral estimation developed by \cite{Wieczorek2005, Wieczorek2007}, \cite{Simons2006}, and \cite{Simons20062} that is adopted in our study. A detailed analysis comparing several methods of spectral localization can be found in \cite{Dahlen2008}. Moreover, theoretical loading models of the lithosphere that consider both surface and subsurface loads have been developed \citep{mcgovern2002, Belleguic2005, Grott2012, Broquet2019}. And lastly, there are improved gravity and topography models \citep{Konopliv1999, Rappaport1999, Wieczorek2015} that were not available for many of the earliests studies. For these reasons, in this study, we reassess the compensation state of the highland plateaus, and attempt to place constraints on their average crustal thickness and elastic thickness. We make use of a flexural loading model and a localized spectral admittance modeling technique employed in many recent studies of Mars \citep{Belleguic2005, WIECZOREK2008, Grott2012, Beuthe2012, Broquet2019, Broquet2020}.

\begin{figure}[h]
  \centering
	\includegraphics[width=\textwidth]{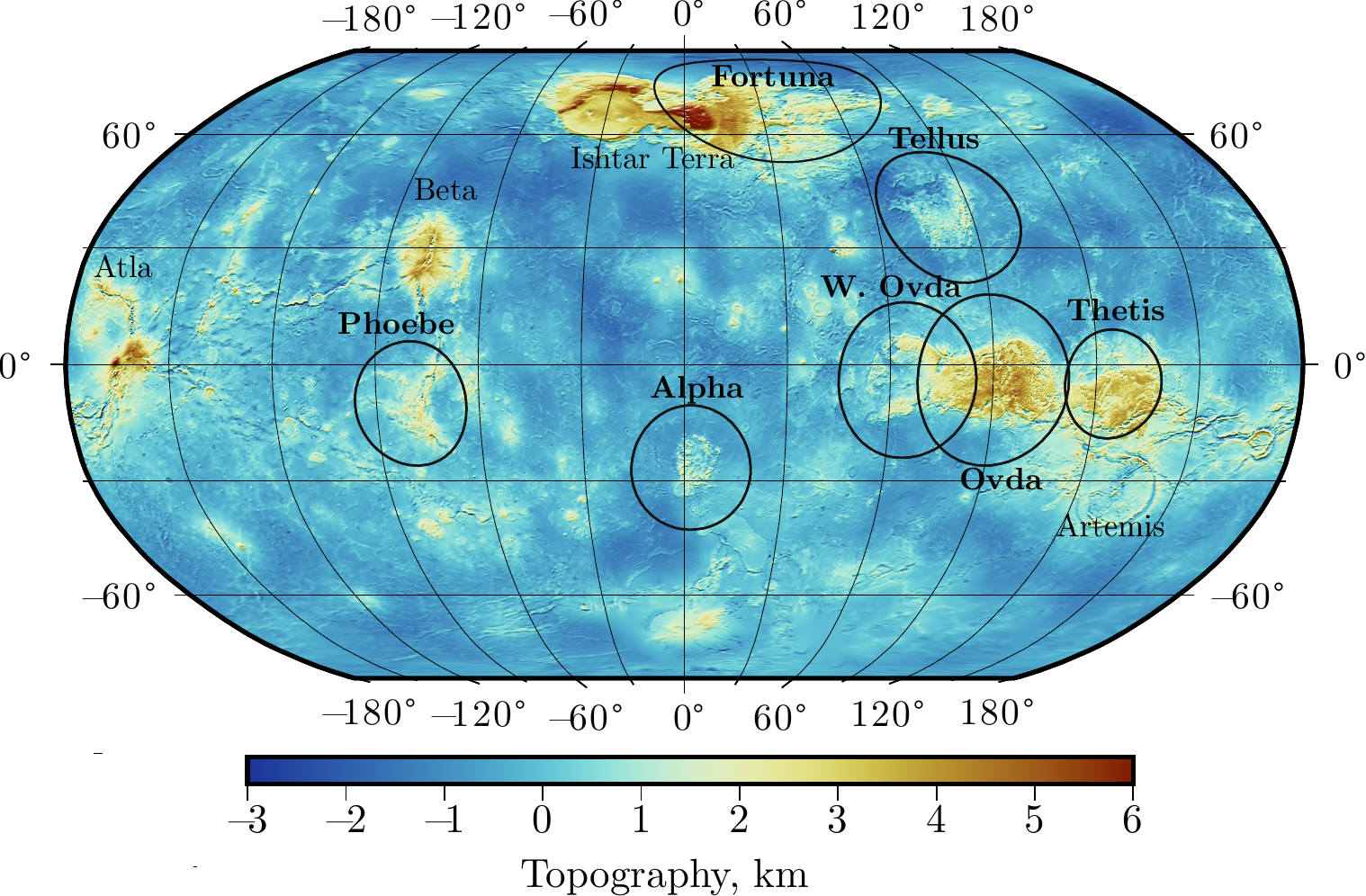}
	\caption{Topography of Venus from the spherical harmonic model VenusTopo719 \citep{Wieczorek2015}, referenced to the mean planetary radius. Major geological structures are annotated and the circled features are the regions we investigate, where the circle size represents the size of the localization window (see section \ref{sec:localization} for details). The image is presented in a Robinson projection and is overlain by a shaded relief map derived from the topographic model.}
	\label{fig:venustopo_text}
\end{figure}

We investigate six Venusian crustal plateaus, namely Thetis, Ovda, Western Ovda, Alpha, Tellus and Phoebe regiones, which are indicated in Figure \ref{fig:venustopo_text}.  We note that although Phoebe Regio was sometimes described as a crustal plateau \cite[e.g.][]{PhillipsHansen1994, Nunes2004} it has been also defined as a transition between plateaus and volcanic rises \cite[e.g.][]{Phillips1998, Kiefer2003} since it presents some geophysical and structural features consistent with both types of features.

The outline of this paper is as follows. In section \ref{sec:methods} we describe the datasets and review the modeling and analysis techniques adopted in our study. We  invert for geophysical parameters, such as crustal thickness and elastic thickness, at every region and present our results in section \ref{sec:results}. In section \ref{sec:discussion} we compare our results with previous studies and discuss the implications of our estimations regarding the compensation mechanism and the thermal evolution of plateaus. Our conclusions are presented in section \ref{sec:conclusion}.

\section{Methods} \label{sec:methods}

\subsection{Gravity and Topography Datasets}\label{sec:gt}

Datasets that are defined on the surface of a sphere, such as the gravity and topography of planets, are commonly analyzed using spherical harmonics, which are the natural set of orthogonal basis functions on the surface of a sphere. A review of the use of spherical harmonics applied to planetary gravity and topography studies can be found in \cite{Wieczorek2015}.

Any square-integrable function, $f$, can be expressed as a linear combination of spherical harmonic functions as

\begin{equation}
    f(\theta, \phi) = \sum_{\ell = 0}^{\infty} \sum_{m = - \ell}^{\ell} f_{\ell m} Y_{\ell m}(\theta, \phi) \mathrm{,}
\end{equation}

where $Y_{\ell m}$ is the spherical harmonic function of degree $\ell$ and order $m$, $ f_{\ell m}$ is the corresponding spherical harmonic expansion coefficient, and $\theta$ and $\phi$ represent the position on the sphere in terms of colatitude and longitude, respectively. Likewise, we are able to express the gravitational potential $U$, at a certain position $\textbf{r}(r,\theta, \phi)$ exterior to the mass distribution of an object, as a sum of spherical harmonic functions

\begin{equation}\label{eq:potential}
    U(\textbf{r}) = \frac{GM}{r} \sum_{\ell = 0}^{\infty} \sum_{m = - \ell}^{\ell} \left( \frac{R_0}{r}\right)^\ell C_{\ell m} Y_{\ell m}(\theta, \phi)\mathrm{,}
\end{equation}

where $G$ is the gravitational constant and $C_{\ell m}$ represents the spherical harmonic coefficients of the gravitational potential at a reference radius $R_0$ for an object with a total mass $M$. In practice, the maximum value of $\ell$ is truncated according to the data resolution. In this study, we use the radial component of the gravitational acceleration that is derived by taking the first partial derivative with respect to the radial component of equation \ref{eq:potential}:

\begin{equation}
    g_r(\textit{r}) = \frac{GM}{r^2} \sum_{\ell = 0}^{\infty} \sum_{m = - \ell}^{\ell} \left( \frac{R_0}{r}\right)^\ell (\ell + 1) \ C_{\ell m} Y_{\ell m}(\theta, \phi)\mathrm{.}
\end{equation}

We note that for all of our analyses we make use of 4-$\pi$ normalized spherical harmonics that are commonly used in geodesy, and also use the sign convention where gravity is positive when directed downwards.

Our analysis approach makes use of the relationship between gravity and topography data in the spectral domain. For this purpose, we derive the power spectrum, which represents how the power of the function varies with spherical harmonic degree. Using the orthogonality properties of spherical harmonics and adopting a generalization of Parseval's theorem it can be shown that the total power of a function $f$ is related to its spectral coefficients by

\begin{equation}
    \frac{1}{4 \pi} \int_{\theta, \phi} [f(\theta, \phi)]^2 \sin{\theta} d\theta d\phi = \sum_{\ell = 0}^\infty S_{\textit{ff}}(\ell)\mathrm{,}
\end{equation}

where

\begin{equation}
    S_{ff}(\ell) = \sum_{m = - \ell}^\ell f_{\ell m}^2
\end{equation}

is the power spectrum of $f$. Similarly, the cross-power spectrum of two functions $f$ and $k$ is defined as

\begin{equation}
    S_{fk}(\ell) = \sum_{m = - \ell}^\ell f_{\ell m}k_{\ell m}\mathrm{.}
\end{equation}

In order to quantify the relation between the amplitudes of gravity and topography signals we use the spectral admittance function

\begin{equation}\label{eq:admittance}
	Z(\ell) = \frac{S_{hg}(\ell)}{S_{hh}(\ell)}\mathrm{,}
\end{equation}

where $S_{hg}$ is the cross-power spectrum of radial gravity and topography and $S_{hh}$ is the power spectrum of the topography. When plotteing the admittance we will use units of mGal km$^{-1}$.

The spectral correlation between gravity and topography, $\gamma$, is a dimensionless parameter involving the power spectra and cross-spectrum of the two functions

\begin{equation}
	\gamma(\ell) = \frac{S_{hg}(\ell)}{\sqrt{S_{hh}(\ell)S_{gg}(\ell)}}\mathrm{,}
\end{equation}

where $S_{gg}$ is the power spectrum of the gravity field. The correlation characterizes the phase relation between the two fields and is bounded between $-1$ and 1. If it is assumed that gravity and topography should be perfectly correlated ($\gamma = 1$), then any correlation value below unity would represent noise  (either measurement noise, or geologic signals not accounted for by the model). In this scenario, it can shown that the admittance uncertainty is

\begin{equation}\label{eq:variance}
	\sigma^2(\ell) = \frac{S_{gg}(\ell)}{S_{hh}(\ell)} \frac{1 - \gamma(\ell)^2}{2\ell}\mathrm{.}
\end{equation}

By expressing the power spectrum of the gravity signal as

\begin{equation}
	S_{gg}(\ell) = S_{gg}^M(\ell) + S_{II}(\ell)\mathrm{,}
\end{equation}

with $S_{gg}^M(\ell)$ and $S_{II}(\ell)$ representing respectively the theoretical gravity without any noise and noise power spectra, we can define the ``signal-to-noise ratio'' as $S_{gg}^M/S_{II}$. Accordingly, we are able to estimate that a correlation value of $0.71$ corresponds to a signal-to-noise ratio of 1 \cite[e.g.,][]{WIECZOREK2008}. This correlation value will be used as a threshold in our admittance analysis.

The highest resolution gravity and topography models of Venus are mainly based on data obtained by the Magellan spacecraft that orbited the planet from 1990 to 1994. We adopt the 719th degree and order spherical harmonic topography model VenusTopo719 \citep{Wieczorek2015}. This model makes use of the Magellan topographic data presented in \cite{Rappaport1999} that covers $\sim 98 \%$ of the surface along with Pioneer Venus and Venera 15/16 topography data that were used to help fill some of the gaps in the Magellan data products. As for gravity, the highest resolution dataset available is the 180th degree and order spherical harmonic solution of \cite{Konopliv1999} named MGNP180U, which is the model adopted in our study. The gravity measurements used for this model come mainly from Doppler tracking of the Magellan mission and are supplemented by earlier data from the Pioneer Venus Orbiter.

In Figure S1 we present the power spectrum of both topography and the radial component of the gravity data, along with the global spectral admittance and correlation. This figure shows that the power of the noise exceeds that of the data around degree 75. In practice, this value is generally used as the global maximum resolution of the gravity model which corresponds to a half-wavelength of approximately 250 km. Nevertheless, we note that the resolution of the MGNP180U model actually varies considerably depending on the location. This is because the altitude of Magellan in the radio tracking phase, after orbit circularization, varied from 600 km at apoapsis to 155 km at periapsis. In addition, there is a gap in low-altitude tracking between longitudes 140--220$^{\circ}$E \citep{Sjogren1997}. The dataset resolution can be examined using the so-called degree strength, which is the spherical harmonic degree where the noise in the data exceeds the signal at a particular latitude and longitude. The degree strength map of MGNP180U \cite[see Figure 3 in]{Konopliv1999} shows that the resolution is predominantly latitude dependent due to the spacecraft altitude although there are important longitudinal variations as well, related to coverage gaps. The degree strength ranges from $\sim$100 near the equator decreasing to 40 at higher latitudes.

\subsection{Lithospheric Compensation Model}\label{sec:litmodel}

We investigate the relation between gravity and topography using a model of the lithosphere to generate theoretical gravity models and compare them to the observed gravity. In order to determine the spherical harmonic coefficients of the predicted gravity $g_{\ell m}$ we presume the following relation:

\begin{equation} \label{eq:gtrel}
	g_{\ell m} = Q_\ell h_{\ell m} + I_{\ell m}
\end{equation}

where $h_{\ell m}$ are the observed topography coefficients, $Q_\ell$ is the model-dependent linear transfer function, assumed to be isotropic (i.e. only dependent of degree $\ell$) and $I_{\ell m}$ is the part of the gravity signal not predicted by the model (such as noise). In the case where  $I_{\ell m}$ is a zero-mean random variable, after multiplying both sides of eq. \ref{eq:gtrel} by $h_{\ell m}$, summing over all orders $m$ and taking the expectation, we find that the transfer function $Q_\ell$ corresponds to the spectral admittance (eq. \ref{eq:admittance}).

In this study we consider that the lithosphere behaves as a thin elastic shell that overlies a fluid interior. The main assumptions and theoretical development of thin elastic shells is presented in \cite{Kraus1967} and explored in the context of topographic loads on elastic lithospheres by \cite{Turcotte1981} with further developments presented in \cite{Beuthe2008}. This formulation establishes the relation  between the pressure of loads acting on the elastic lithosphere to the vertical deflection of the surface. The amount of deflection is controlled by the flexural rigidity of the lithosphere which depends on three parameters: Poisson's ratio $\nu$, Young's modulus $E$, and the elastic thickness $T_e$.  The lower the elastic thickness the more flexible is the lithosphere, which increases the lithospheric deflection. The Airy isostasy regime, in which the topography is locally compensated, corresponds to the end-member case where $T_e = 0$ km. Since the elastic thickness is related to the temperature of the lithosphere, this parameter can provide insights about the lithosphere's thermal structure, as discussed later in section \ref{sec:disheatflow}. In this study we fix Poisson's ratio and Young's modulus as constants, with $\nu=0.25$ and $E=100$ MPa. These values are based on what is known for terestrial rocks and are consistent with what has been used in previous studies of Venus.

We adopt the loading model of \cite{Broquet2019} that allows for loads emplaced both onto and below the surface. This model is comparable to previous loading models developed for Mars by \cite{mcgovern2002}, \cite{Belleguic2005}, \cite{Beuthe2012} and \cite{Grott2012}. Surface loads can be the result of extrusive volcanism or topographic highs generated by tectonic processes. In the case of subsurface loads, we consider two possible scenarios: either as a low-density layer in the mantle, such as from a buoyant mantle plume or depleted layer, or a high density layer within the crust, which could correspond to dense magmatic intrusions. The subsurface loads are emplaced at depths $z=150$ km for the mantle plume scenario and $z=15$ km in the case of intrusions and both are modeled as thin mass-sheets with negative and positive density anomalies with respect to their surroundings. The mantle plume depth is based on inferences of the thermal boundary layer depth \citep{simons_1997}. Nevertheless, 	after testing a range of load depth values we observed that this parameter does not significantly affect our results. We note that the model of dense intrusions in the crust could be used as a proxy for a model with a dense mantle downwelling beneath the crustal plateaus.

In our modeling we consider that the surface and subsurface loads are in-phase and that the ratio of subsurface and surface loads is isotropic and independent of wavelength, which is a valid assumption when the correlation is high. Thus, we can parametrize the ratio of subsurface and surface loads by

\begin{equation}
	 f = \frac{\Delta \rho_z dz}{\rho_l(h - w)}
\end{equation}

where $\Delta \rho_z$ is the density difference between the internal load and its surroundings, $dz$ is the thickness of the sursurface load, $\rho_l$ is the density of the surface load, $h$ is the topography and $w$ is the lithospheric deflection. In order to have a bounded loading parameter we define the load ratio variable as $L = f/(|f|+1)$ that can vary between $-1$ and $1$. In this parametrization, $L=0$ represents the presence of surface loads only, while $L=-1$ and $L=1$ correspond respectively to pure bottom-loading with low-density (mantle plume scenario) and high-density (dense intrusions scenario) mass-sheets. Although the surface load could have a density different from the crust, in this study we consider that the density of the two are the same. 

At this point, we remark that our loading model differs somewhat from the model of \cite{Forsyth1985} that was previously applied to Venus \citep{Phillips1994, Smrekar1994, smrekar2003, Hoogenboom2004, Hoogenboom2005}. The fundamental difference between this model and ours concerns how subsurface loads are treated. In \cite{Forsyth1985}, the subsurface loads are assumed to be on average uncorrelated with the surface loads, whereas in our model the loads are assumed to be perfectly correlated. Though he acknowledged that tectonic and volcanic processes would likely favor the formation of correlated loads, it was argued that erosion of the continental crust on Earth could act to decorrelate surface and subsurface loads. Since there is little to no erosion occurring on Venus today, the assumption of uncorrelated loads for this planet is thus suspect. Regardless, it is important to emphasize that the differences between the two models become less prominent as the elastic thickness decreases, and that the two models are in fact identical for the case of Airy isostasy (i.e., $T_e=0$). A second difference is that in Forsyth's original approach, the subsurface load was determined from the observed gravity field, and because of this, the admittance function was fit perfectly. Their analysis instead focussed on interpreting the correlation function. A benefit of this approach is that the loading ratio $f$ is allowed to vary for each wavelength. Nevertheless, it should be noted that other formulations that make use of Forsyth's assumption of uncorrelated loads exist. For example, \cite{Phillips1994} assumed a constant loading ratio $f$ and modeled both the admittance and correlation function in a manner that is very similar to our approach

In addition to the parameters formerly described, the transfer function, $Q_\ell$, depends on the thickness of the crust $T_c$, crustal density $\rho_c$, mantle density $\rho_m$, and the planetary radius at the localized region $R_{loc}$. The latter is determined by computing the ratio of the degree and order zero terms of the localized topography and window function. The full derivation and final transfer function equation under the mass-sheet approximation can be found in Appendix B of \cite{Broquet2019}. For our purposes, it is sufficient to write it schematically as

\begin{equation}\label{eq:f}
	 Q_\ell = Q_\ell(\ell,  T_e, T_c, \rho_l, \rho_c, \rho_m, L, z, R_{loc}, E, \nu)\textrm{.}
\end{equation}

Our investigation will consider three free parameters: the elastic thickness $T_e$, crustal thickness $T_c$ and load ratio $L$. All the other parameters will be set to constants and the values used are compiled in Table \ref{tab:parspace}. Although it would be useful to estimate the crustal density of the highland plateaus we were not able to constrain this parameter due to the low resolution of the available gravity field (see Figure S2).

With the above flexural loading model of the lithosphere, we can calculate the predicted gravitational signal for a given topography and the modeled admittance. Since most features on Venus have relative small topographic reliefs, with the exception of Maxwell Montes located at Ishtar Terra, in this study we make use of the mass-sheet (first order) approximation when computing the predicted gravity from the relief along a density interface. In the regions of interest the difference between the mass-sheet and higher-order approximations does not exceed 2 mGal.
\begin{table}[h]
\caption{Parameter values consider in our inversions.}
\centering
\begin{tabular}{l l}
\hline
Parameter & Value  \\
\hline
Elastic thickness $T_e$ & 0 -- 75 km  \\
Crustal thickness $T_c$ & 1 -- 100 km  \\
Load ratio $L$ & $-0.8$ to $0.8$ \\
Mantle density $\rho_m$ & 3300 kg m$^{-3}$  \\
Crustal density $\rho_c$ & 2800 kg m$^{-3}$  \\
Load density $\rho_l$ & 2800 kg m$^{-3}$  \\
Load depth $z$ & 150 ($L<0$) or 15 ($L>0$) km\\
Young's modulus $E$ & 100 GPa  \\
Poisson's ratio $\nu$ & 0.25 \\
\hline
\label{tab:parspace}
\end{tabular}
\end{table}

\begin{figure}[p]
  \centering
	\includegraphics[width=\textwidth]{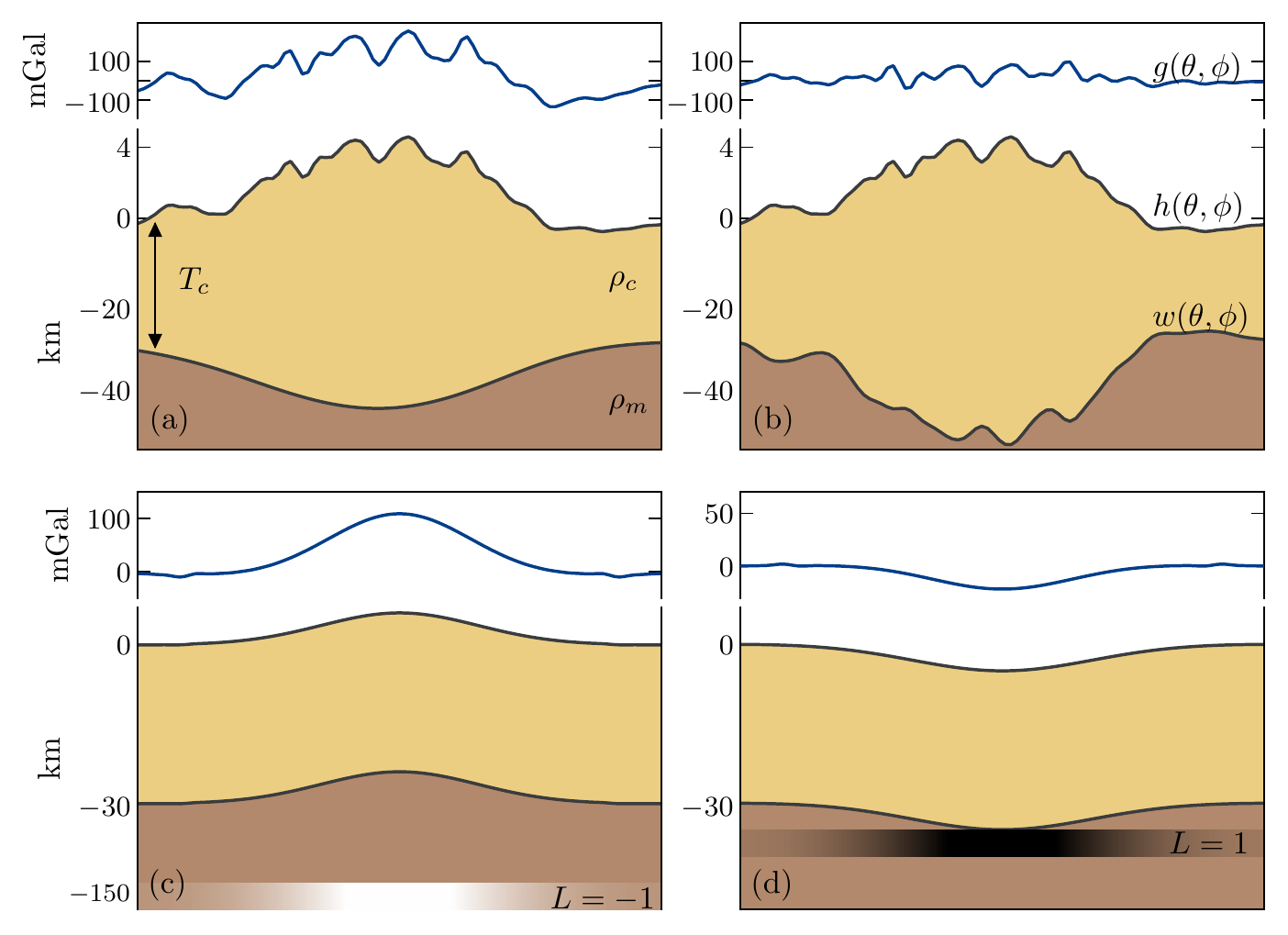}
	\caption{Schematic view of the adopted flexural model, showing the topography profile $h(\theta,\phi)$ and associated deflection $w(\theta,\phi)$. The corresponding radial gravity signal, $g(\theta, \phi)$, is represented by the blue curve at the top of each subplot. Panel (a) shows the case of surface loads with $T_e$=300 km while (b) represents surface loading with $T_e$=15 km. Panel (c) displays the subsurface loading scenario in which the topography is supported by a buoyant layer in the mantle, represented by the thin layer at 150 km depth, with light colors indicating a negative density anomaly and (d) exemplifies the  case of a high-density layer, such as intrusions, represented by the thin layer at the base of the crust with dark colors indicating positive density anomalies. Panels (c) and (d) have a 15 km elastic thickness and the vertical scale of both surface relief and the crust-mantle interface are doubled for visual purposes. All profiles measure approximately 4000 km across. The models have a crustal thickness of 30 km, crustal and load densities of 2800 kg m$^{-3}$ and a mantle density of 3300 kg m$^{-3}$.}
	\label{fig:modelscheme}
\end{figure}

In Figure \ref{fig:modelscheme} we present an illustration of the possible loading scenarios and their gravitational signatures. Panels (a) and (b) represent the surface loading model (also known as top-loading) for elastic thicknesses of 300 km and 15 km, respectively. We can see a considerable difference in the gravity for these two cases, with a more rigid lithosphere being associated with a stronger signal while for a flexible lithosphere the longer wavelengths are nearly completely compensated and the gravity signals are related to the shorter-wavelength relief, which are uncompensated. Panel (c) shows the scenario of topographic support by a buoyant mantle layer, commonly referred to as a bottom-loading model, and (d) illustrates the flexure caused  by high-density subsurface loads at the base of the crust. In all examples we can clearly see that gravity and topography are perfectly correlated. In fact our model is constructed in such a way that the global spectral correlation is always equal to 1 or $-1$ (perfectly correlated or anti-correlated).

\subsection{Localized Spectral Admittance Analysis}\label{sec:localization}

The previous sections showed how one can analyze the spectral relation between gravity and topography in a global perspective. However, planets such as Venus present an important diversity of geologic features associated with distinct evolutionary paths. Thus, we must employ an analysis technique that allows us to extract information from data that are localized to specific regions of interest. Some techniques analyze such data using purely spatial domain approaches, whereas others use more sophisticated spatio-spectral localization techniques.

During the Magellan era, many gravity studies used geoid-to-topography ratios (GTRs) to investigate the interiors of various regions on Venus \cite[e.g.,][]{smrekar1991, kuncinskas_1994, moore_1997}. GTRs provide, for a finite region, a single number that characterizes the best fitting linear relation between the geoid and topography. As a purely spatial method, this technique preserves precise spatial information but loses all spectral information, including a potential wavelength dependency on the investigated parameter. A shortcoming of this technique is that it is not easy to remove long-wavelength gravity signals that are comparable to or larger in size than the analysis region.  It is also more difficult to validate the assumptions of the theoretical model, such as whether the region is Airy compensated or not.

Meanwhile, some of the first localized admittance studies were performed for Venusian regions. \cite{grimm_1994} applied Cartesian spatio-spectral windowing functions to investigate crustal plateaus. Following this, \cite{simons_1997} introduced a spherical wavelet-like analysis to perform admittance studies over various features on Venus. This wavelet technique was later used in several investigations on the planet \cite[e.g.][]{smrekar2003, Hoogenboom2004, anderson_2006}. In this technique, the size of the analysis region is wavelength dependent, and scales with the size of the wavelet. Spatio-spectral localization techniques are subject to a well known compromise between spatial and spectral resolution, and considering the generally low resolution of planetary gravity datasets, these methods are usually restricted to large features.

In our study, the localization technique introduced by \cite{Wieczorek2005, Wieczorek2007} is adopted. In this approach, the localization is performed by multiplying the gravity and topography data by a fixed localization window and expanding the results in spherical harmonics, in order to retrieve the localized spectral estimate for a specific region. The spatially localized field $F(\theta, \phi)$ of a global function $f(\theta, \phi)$ is given by

\begin{equation}
	F(\theta, \phi) = w(\theta, \phi) f(\theta, \phi)\mathrm{,}
\end{equation}

where $w$ is the localization window. The main difference between this and the wavelet-based techniques is that the window is fixed in size and geometry for all wavelengths. This allows the analyst to ensure that all signals are from a single geologic region of interest.

In the spectral domain the window power spectrum and the global function spectrum are related by an operation that resembles a convolution \cite[for details see Appendix B of]{Wieczorek2007}. Accordingly, each degree $\ell$ of the localized power spectrum contains contributions from the global power spectrum within the range $\ell \pm \ell_{win}$, where $\ell_{win}$ is the spectral bandwidth of the window. Thus, $\ell_{win}$ should be as small as possible to minimize smoothing in the spectral domain and to maximize the number of uncorrelated spectral estimates. On the other hand, narrowing the bandwidth reduces the spatial concentration of the window, i.e., a larger portion of the window power would be outside the region of interest. For a given maximum spectral bandwidth $\ell_{win}$, our windows are designed to optimally concentrate the power of the window within the region of interest. Though this technique provides several orthogonal windows, given the low spatial resolution of the Venusian gravity field, we only make use of the first, best concentrated, window.

We make use of the Python pyshtools package \citep{Wieczorek2018} to generate localization windows that are concentrated within a spherical cap. The necessary inputs are the angular radius, $\theta_0$, of the cap and the window bandwidth $\ell_{win}$. In practice, for a specified angular radius, we tune the window bandwidth $\ell_{win}$ such that 99\% of the window power is concentrated in the region being studied. We notice that because the window has a bell-like shape \cite[e.g., Figure 1 of]{Wieczorek2005}, the signal close to the edge of the cap is strongly attenuated. In fact, for a 99\% concentration window with angular radius $\theta_0$, about 50\% of the power is contained within a radius of approximately $\theta_0/2$. Because of this property we design our windows to be somewhat larger than the feature being analyzed.

\section{Results}\label{sec:results}
In the first part of this section the modeling and inversion procedures are described in detail for Alpha Regio. Following this, we summarize the inversion results for the other five investigated crustal plateaus: Ovda, Western Ovda, Thetis, Tellus and Phoebe regiones. We note that the only major tessera region that has not been included in our analysis is Fortura Tessera, located in the eastern part of Ishtar Terra. The observed localized correlation in this region is very low, mostly lower than the adopted threshold of $0.71$ and always lower than $0.75$. It is not obvious if this attribute is due to uncorrelated gravity signals coming from the deep crust and mantle or poor data quality in the region.

\subsection{Case Study: Alpha Regio}\label{sec:resalpha}
\begin{figure}[h]
  \centering
	\includegraphics[width=\linewidth]{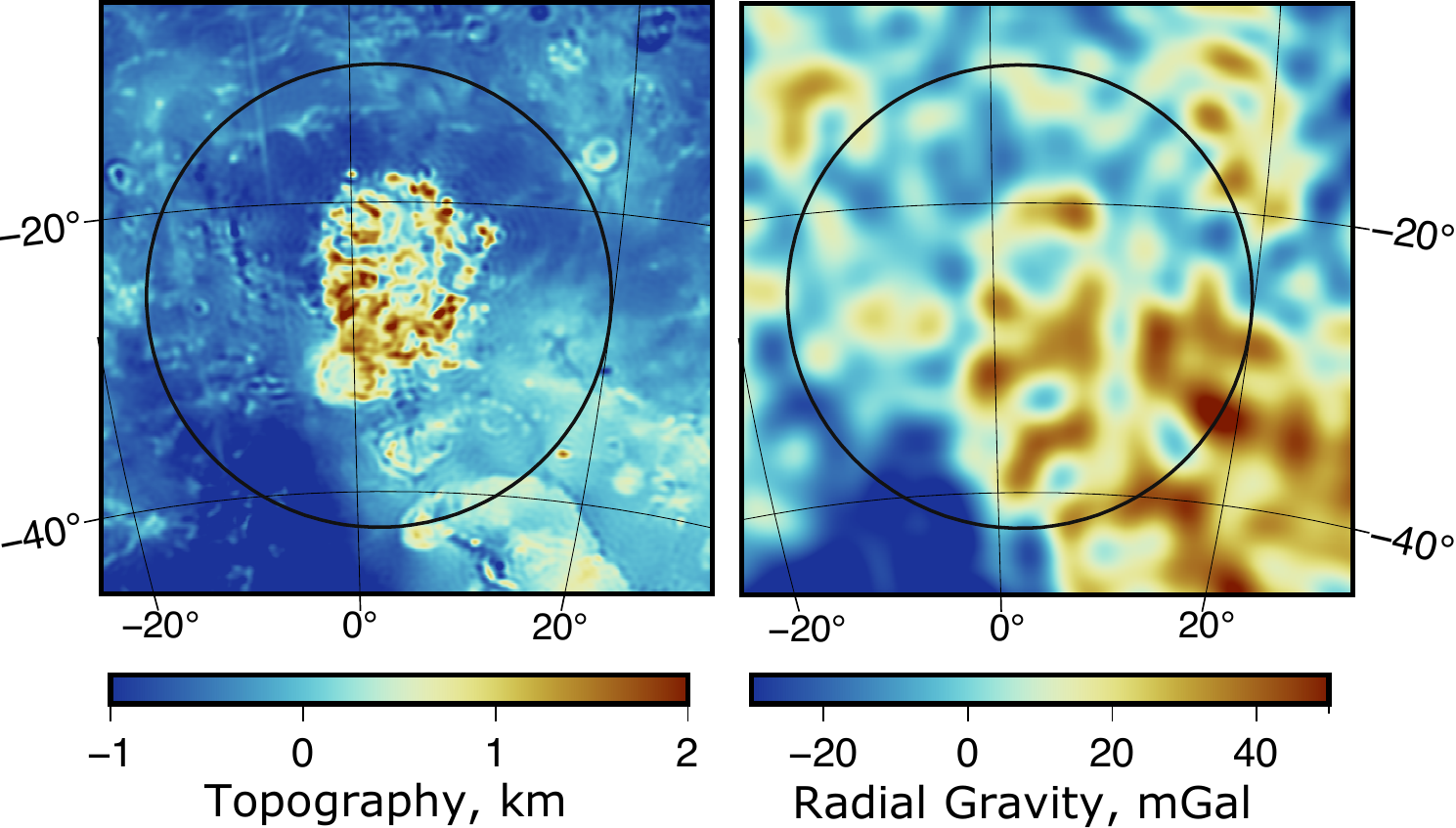}
	\caption{(left) Topography from VenusTopo719 and (right) radial free-air gravity anomaly from MGNP180U after truncating spherical harmonic degrees above 100 at Alpha Regio. The circles correspond to the limits of the adopted localization window, with $\theta_0 = 16^{\circ}$ corresponding to a diameter of 3380 km. The adopted projection for both maps is Lambert azimuthal equal-area.}
	\label{fig:alpha_gridtopograv}
\end{figure}

Alpha Regio is an $\sim$1300 km wide radar-bright crustal plateau located at latitude 25$^\circ$S  and longitude 2$^\circ$E. This isolated plateau is characterized by tessera terrain and steep-sized topography, standing on average 1 km above the surrounding volcanic plains. Immediately south of Alpha is located the 330 km diameter Eve Corona that marks the 0$^\circ$ meridian of Venus. Detailed geologic descriptions of Alpha Regio can be found in \cite{Bindschadler1992} and \cite{Bender2000}. Figure \ref{fig:alpha_gridtopograv} shows the topography and radial gravity at this region where we can see positive gravity anomalies related to the high topography of the plateau. Though some of the gravity anomalies are clearly correlated with the high-standing topography, we note that there exist gravity anomalies of similar amplitude exterior to this plateau that are not strongly correlated with geologic features.

The first step in our analysis is to compute the localized observed spectral admittance and correlation, which is done by applying the windowing function as described in section \ref{sec:localization}. We chose the window size to encompass the entire feature while avoiding gravitational and topographic signals exterior to the plateau. We used a window with $\theta_0=16^\circ$, corresponding to a diameter of 3380 km, which results in $\ell_{win}= 16$ given that a 99\% concentration factor is adopted. Then, we downward continued the observed gravity field to the average local radius and computed the localized admittance and correlation. At Alpha, the average radius corresponds to 6051.86 km, which is very close to the planetary average of 6051.88 km.

We proceed by defining the range of spherical harmonic degrees that will be used to perform the inversion. The upper limit is determined by the resolution of the data at Alpha ($\ell_{loc}$), which we take to be the degree strength from the map of \cite{Konopliv1999}. For Alpha, the degree strength is 75. Since each degree has contributions from $\pm \ell_{win}$ we avoid the inclusion of noise-dominated data in our analysis by removing all degrees above  $\ell_{loc} - \ell_{win}$, resulting in a upper limit of degree 59 for the region. We also ignored all degrees in the localized spectra that were less than $\ell_{win}$ because these are dominated by signals with wavelengths that are greater than the window size and because these degrees typically have extremely high variances \citep{Wieczorek2007}.  Though we never analyze data outside of this degree range, we sometimes avoid degrees where the correlation is smaller than 0.71, corresponding to a signal-to-noise ratio less than 1 (see section \ref{sec:gt}).

\begin{figure}[h]
	\centering
	\includegraphics[width=0.8\linewidth]{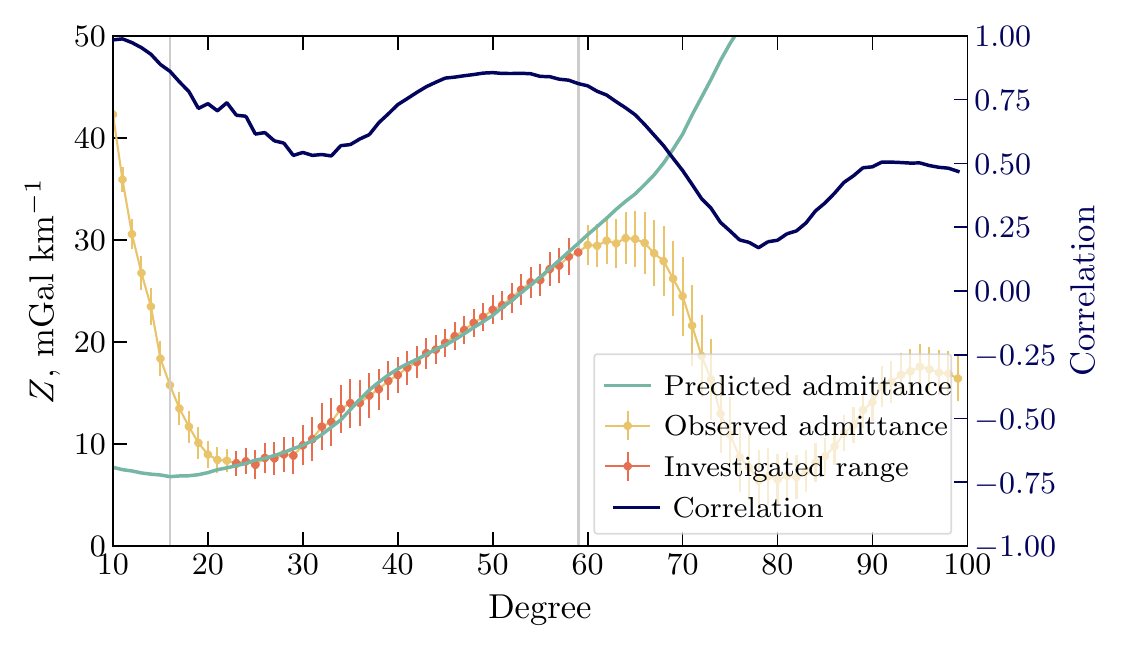}
	\caption{Observed localized spectral admittance (left axis) and correlation (right axis) at Alpha Regio. The points in red represent the spherical harmonic degrees used in our inversion. Grey lines show the theoretical investigation limits $\ell_{win}$ and $\ell_{loc}$-$\ell_{win}$. The green curve shows the best-fitting admittance, which corresponds to $T_c=15$ km, $T_e=20$ km and $L=0$.}
	\label{fig:alpha_admit}
\end{figure}

The localized admittance, correlation and the investigated degree range for Alpha Regio are shown in Figure \ref{fig:alpha_admit}. The correlation is high between degrees 40 to 59 with implied signal-to-noise ratios greater than 1,  which was the range we initially chose to perform the inversion. The drop in correlation that starts at degrees above 60 is likely due to the strong influence of noise in the data while the mild drop around degree 30 is probably caused by true geologic signals that are not correlated with the topography. Regardless, we noticed that the degree range between 23--40 provided a very good fit to our best-fitting theoretical model. Thus, in order to increase the degrees of freedom and robustness of the inversion, we included these data points in the investigation resulting in an expanded degree range from 23 to 59, as shown in Figure \ref{fig:alpha_admit}. Moreover, although the degrees 17--20 have a high correlation and could in principle be used in the modelling, we found that these degrees never  satisfactorily fit our best-fitting theoretical model. In all likelihood, the high admittances for $\ell < 20$ are the result of processes occurring deep in the mantle \cite[e.g.,][]{Pauer2006}.


In order to determine the best-fitting model parameters and uncertainties, we make use of an exhaustive grid-search of the parameter space. We systematically varied crustal thickness, elastic thickness and load ratio to generate modeled admittances and compared them to the observed admittance over the selected degree range. The model misfit is defined as the root mean square (rms) error:

\begin{equation}
	    \textrm{rms} (T_c, T_e, L) = \sqrt{\frac{1}{N} \sum_{\ell = \ell_{min}}^{\ell_{max}} \left[ Z_{obs}(\ell) - Z(\ell, T_c, T_e) \right]^2}
\end{equation}

where $Z_{obs}$ is the observed admittance, $Z$ is the localized admittance predicted by the model and $N$ is the number of degrees adopted in the summation. The best fitting curve for Alpha is shown in Figure \ref{fig:alpha_admit}. Once the entire misfit function is computed we are able to determine the accepted range of investigated parameters by the use of a maximum allowable misfit threshold. Here we used a 1-standard-deviation criterion based on the average uncertainty of the observed localized admittance

\begin{equation}
    \bar{\varsigma} = \sqrt{\frac{1}{N} \sum_{\ell = \ell_{min}}^{\ell_{max}} \varsigma^2(\ell)}
\end{equation}

with $\varsigma^2(\ell)$ representing the localized version of $\sigma^2$, defined by equation \ref{eq:variance}. This criterion has been commonly used in Martian studies \citep{WIECZOREK2008, Grott2012, Broquet2019}. In previous gravity studies on Venus, other workers have chosen more flexible, but arbitrary, criteria, such as 1.5 times the minimum rms \cite[e.g.,][]{smrekar2003, Hoogenboom2004}. Other approaches for defining a maximum allowable misfit are discussed in \cite{Broquet2019}.

\begin{figure}[h]
  \centering
	\includegraphics[width=\linewidth]{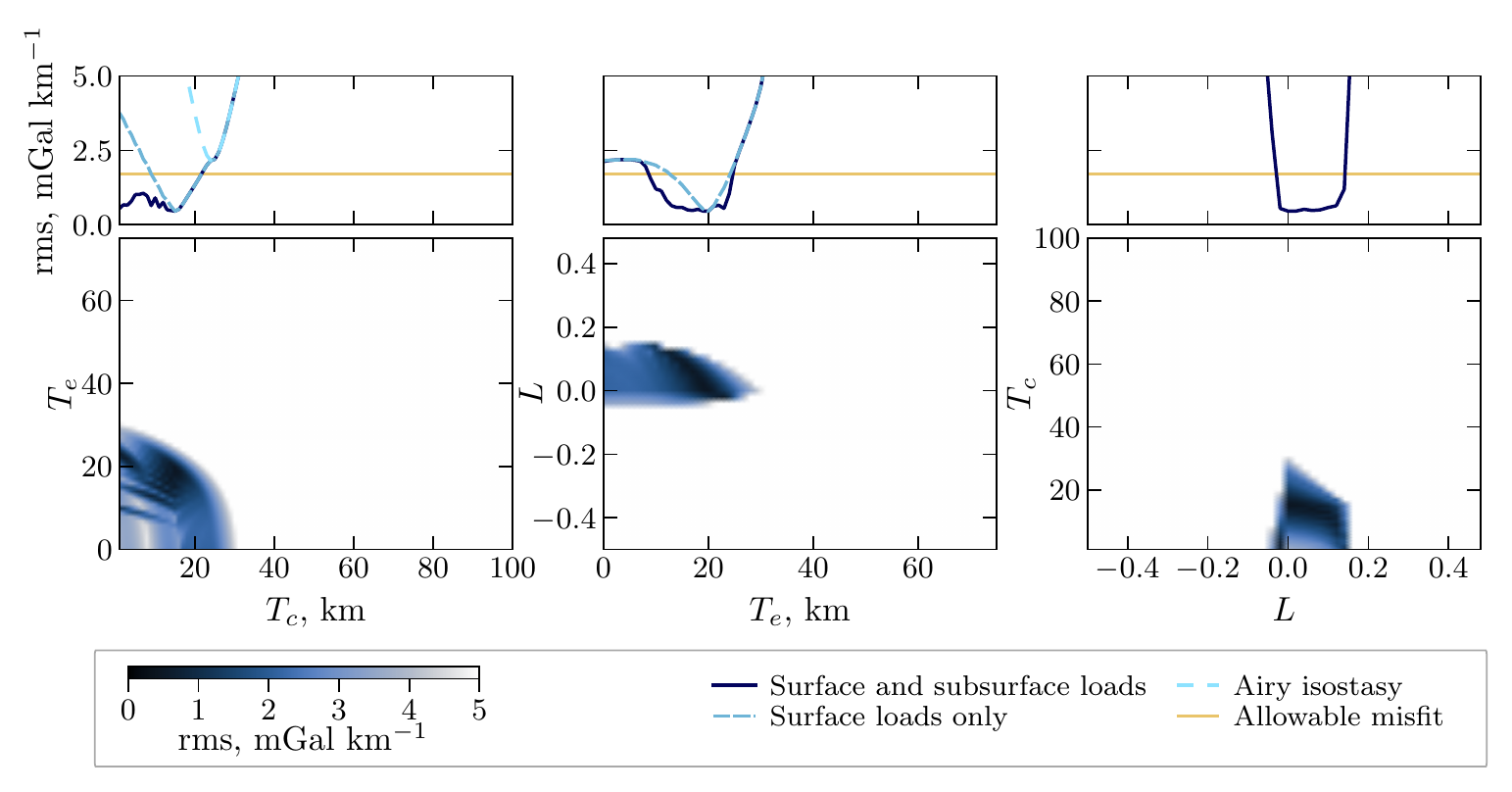}
	\caption{One-dimensional (upper) and two dimensional (lower) misfits between model and observation at Alpha Regio as a function of the crustal thickness, elastic thickness and load ratio. The 1-dimensional misfit plots show the minimum misfits for three different loading scenarios, where the dark blue curves represent the case where surface and subsurface loads are allowed (3 free parameters), the lighter blue curves shows the case where only surface loads are present (i.e., $L=0$) and the cyan curve in the upper left plot indicates an Airy isostasy regime scenario ($T_e = 0$). The 2-dimensional plots in the bottom row display the minimum misfits in terms of two free parameters for our full model with subsurface loads. All two-dimensional plots share the same color scale and the one-dimensional plots share the same $x$ axis as the underlying two-dimensional plot.}
	\label{fig:alpha_misfit}
\end{figure}

In Figure \ref{fig:alpha_misfit} we plot various representations of the misfit functions for Alpha Regio. The three upper plots display 1-dimensional minimum misfit curves for the crustal thickness, elastic thickness and load ratio. Looking at the dark blue curve  (which is for our full inversion that includes both surface and subsurface loads), in the left plot we find that the crustal thickness is constrained only to be less than 21 km. In the center plot, we find that the elastic thickness is constrained to lie between 9 and 24 km. In the right plot, the load ratio misfits shows that this parameter is essentially constrained to the presence of surface loads only ($L=0$) or small positive loads, indicating a potential small dense crustal intrusion, where $L=0$ is associated with the best-fitting value.

Given that the load ratio is consistent with being zero, we decided to test two simpler scenarios: one having only surface loads ($T_c$ and $T_e$ as free parameters) and a second that is in an Airy isostasy regime ($T_c$ being the only free parameter). The misfit curves for the first scenario ($L=0$) is overplotted in the upper two figures in light blue. We find that when subsurface loads are neglected this does not affect the upper limit for the crustal thickness, and that we are able to obtain a firm lower bound on the crust thickness of 9 km. Similarly, for the elastic thickness, we find that this does not affect our upper bound of 24 km, but that it slightly reduces the range of the lower bound to 12 km. The case where we assume Airy isostasy is plotted in cyan in the upper left plot. Here, we find that the best fitting model has a misfit that is greater than our cutoff value. This implies that the assumption of Airy isostasy is not appropriate for this region, and that if this were assumed, the crustal thickness would be biased towards larger values.

In the three bottom plots, we show the minimum misfit as a function of two variables, which allows us to evaluate correlations and trade-offs between parameters. The misfit plot considering crustal and elastic thicknesses (left panel) is the most complex, presenting several local minima and degeneracies. However, there is the expected trend of decreasing elastic thickness with increasing crustal thickness which is related to the attenuation of the compensating signal from the crustal roots. In the middle panel, we can see that there is a trade-off between elastic thickness and the load ratio, where the addition of positive loads decreases the elastic thickness. Regarding the relation between crustal thickness and load ratio (right panel), the addition of positive loads allows for moderately lower crustal thickness. In this situation, the addition of slightly denser material within the crust essentially counterbalances the reduction of crustal material associated to a thinner crust. On the other hand, the addition of small buoyant loads in the mantle leads to crustal thickness values lower than 10 km.

In summary, when subsurface loads are allowed, our best fitting values and 1-sigma limits of the inversion parameters are the following: $L=0(-0.02,0.14)$, $T_e=20(9,24)$ km and $T_c=15(1,21)$\,km. The setting with surface loads only results in $T_e=20(12,24)$\,km and $T_c=15(9,21)$\,km. Overall, these two models result in very similar parameter estimation. The most remarkable difference is that when subsurface loads are not considered, lower and upper bounds are obtained for the crustal thickness, whereas when subsurface loads are considered, only an upper bound is possible. Finally, for the Airy isostasy regime, there are no crustal thickness values that result in acceptable fits for Alpha Regio, the lowest rms value is given by $T_c=24$\,km, which is higher than the values that consider elastic support. We should remark that we do not investigate the possibility of topographic support by Pratt isostasy, i.e. by lateral variations of crustal density between the plateaus and the volcanic plains. In fact, our analysis approach is not well-suited to investigate this compensation regime, since the use of localization windows supress most of the signal coming from the plains. In any case, it is unlikely that the entirety of the plateaus are purely Pratt compensated since this would imply portions to have have extremely low densities. For example, given that the plateaus can reach up to 4 km altitude, assuming that the crustal thickness at the mean planetary radius is of 20 km and that the volcanic plains have a density of 2800 kg m$^{-3}$, the highest elevations of the plateaus would have a density of 2300 kg m$^{-3}$, which is considerably lower than the density of quartz. Moreover, previous studies that compared the two isostatic support mechanisms at crustal plateaus tended to favor Airy over Pratt isostasy \citep{kuncinskas_1994}.

In order to limit the parameter space of our inversions several model parameters were fixed to constant values, as indicated on Table \ref{tab:parspace}. One of these parameters is the crustal density, here assumed to be 2800 kg m$^{-3}$ which is consistent with basaltic rocks on Earth. However, new studies of the emissivity signature of Alpha Regio suggest that the composition of this plateau could be more felsic than the surrounding plains \citep{Gilmore2015}. Therefore, we also performed inversions considering a crustal density of 2650 kg m$^{-3}$ which is a standard value for the density of terrestrial granites. The main effect of this modification was an overall increase of 2 km in the estimated crustal thickness values for all investigated regions. Moreover, changing the depths of the internal loads also have a minor effect on the inversion results. For example, increasing the load depths to 30 and 300 km for the positive and negative loads, respectively, had overall no impact on the elastic thickness estimations and increased the upper limit of accepted crustal thickness values by a few kilometers.

Sensitivity analyses were also performed regarding the size of the localization window and the admittance spectral range used for the inversions, since the choice of these parameters is somewhat subjective. We did inversions considering a window that was 30\% larger and also changed the investigated degree range to 40--59, which corresponds to the degrees with the highest spectral correlations (Figure \ref{fig:alpha_admit}). The change in the degree range had only a small impact on the values of the estimated parameters. The best-fitting values and uncertainties obtained varied within a range of 15\% with respect to the results presented above. Increasing the size of the localization window had a somewhat larger effect, in which the best fitting crustal thickness estimations for the top-loading model changed from 15\,km to 21\,km, corresponding to a 40\% increase. This shift is probably caused by the large gravity signal of the surroundings of Alpha that are unrelated to the plateau itself. These sensitivity tests were performed in all studied plateaus and, as for Alpha Regio, most results varied within a range of 15\% relative to the results presented in the text. In all tests we found that the results were consistent with those presented in the text within their respective uncertainties.

\subsection{Crustal Plateaus}

\begin{figure}[p]
  \centering
	\includegraphics[width=0.9\textwidth]{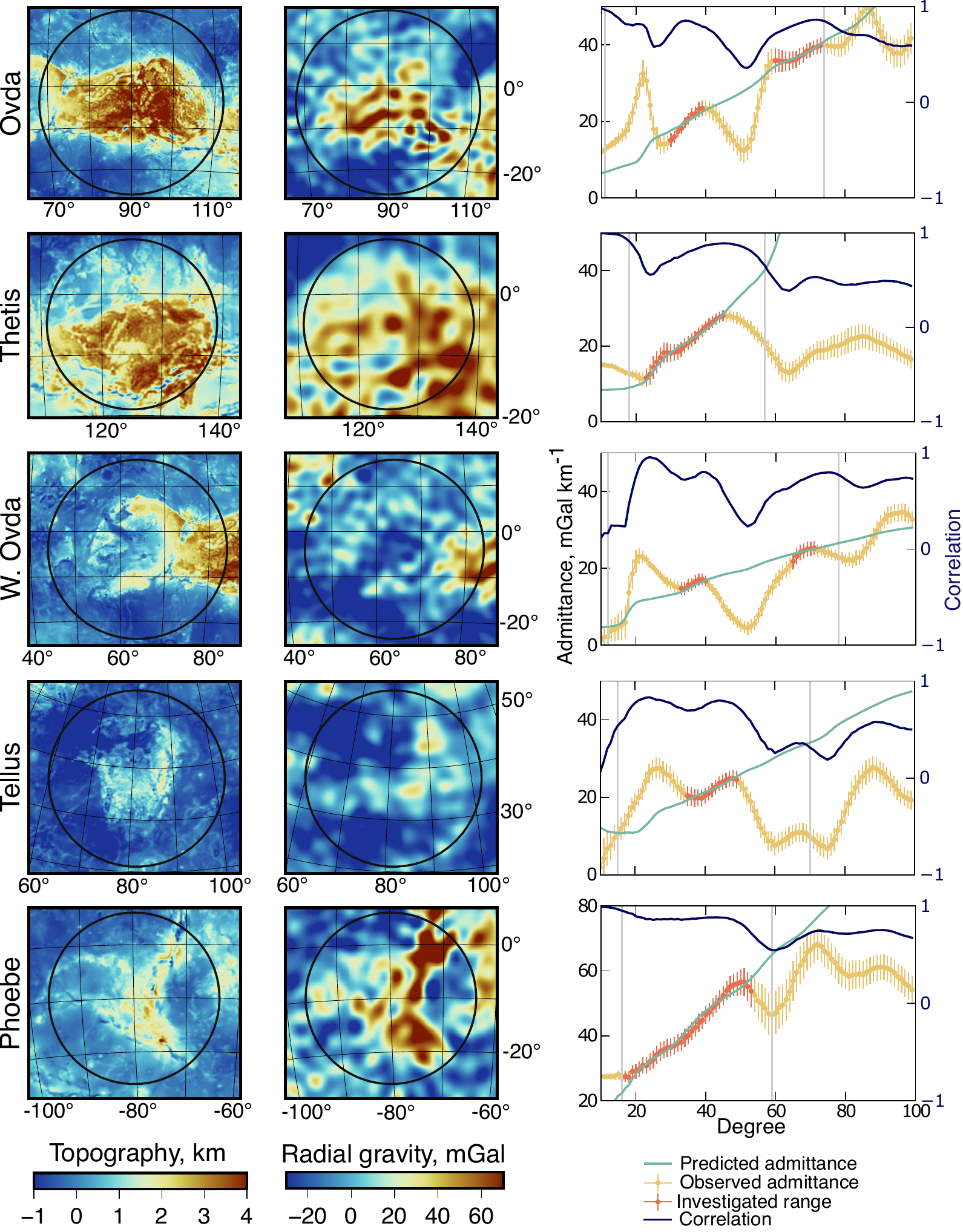}
	\caption{Topography (left) and free-air gravity maps (center) of the 5 investigated crustal plateaus with circles representing the localization window sizes. The adopted projection for these maps is Lambert azimuthal equal-area. (Right) observed admittance, correlation and best-fitting modeled admittance spectra. Grey lines show the maximum possible investigation limits of $\ell_{win}$ and $\ell_{loc}-\ell_{win}$.}
	\label{fig:allreg_obs_win}
\end{figure}

Following the same procedures as presented for Alpha Regio, we estimated the crustal thickness, elastic thickness and load ratio values for the other five crustal plateaus of our study: Ovda, Western Ovda, Thetis, Tellus and Phoebe regiones. Information on the degree range and localization window parameters used for each region is provided in Table S1. In Figure \ref{fig:allreg_obs_win} we present the topography and gravity maps for each of the five plateaus, where the circles represent the sizes of the localization windows we used. In the right column of the figure we show the localized spectral admittance and correlation, and the best fitting theoretical model of the admittance for each region, similar as shown previously in Figure \ref{fig:alpha_admit} for Alpha Regio. Because of the geological complexity of the crustal plateaus, their admittance and correlation spectra are also complex. The correlation is generally not uniform across the entire degree range, and the best fitting model does not adequately fit the observations over the entire investigation range $\ell_{win} < \ell < \ell_{loc} - \ell_{win}$ represented by the vertical lines on the plots.

A common aspect to all five regions is a substantial drop in the localized admittance and correlation spectra around degrees 50, in the case of Ovda and Western Ovda, or 60 for Tellus, Thetis and Phoebe regiones, corresponding to wavelengths of 760 km and 630 km respectively. We systematically disregard those degrees in our analysis where the correlations fall below 0.71, which corresponds to signal-to-noise ratios less than unity. The cause of these reductions in correlation, which are also accompanied by a drop in the admittance,  is not immediately obvious. One possibility is that these are a result of deficiencies in the gravity model that originate from noise and uneven radio tracking coverage. Alternatively, it is possible that these drops in correlation could be real geophysical signals that are plausibly related to processes in the upper mantle, or to processes related to crustal delamination. Lithospheric delamination related to compressive tectonics and crustal thickening was proposed by \cite{Romeo2008} to explain the formation of crustal plateaus and is also predicted by numerical models of coronae formation \citep{Gulcher2020}. In addition to the degree range with low correlations near 50-60, we note that our model also can not account for the observed admittances for degrees less than about 30 for Western Ovda, Ovda and Tellus regiones. In contrast to the degrees near 50-60, these lower degrees are associated with  relatively high correlations, and correspond to wavelengths greater than 1300 km, which are comparable to the size of the plateaus. We suspect that this long wavelength signal is related to processes in the upper mantle that are not adequately accounted for by our loading model that uses a single loading ratio for all spherical harmonic degrees. Since our model was unable to properly fit these components they were removed from the analysis.

\begin{figure}[h!]
  \centering
	\includegraphics[width=0.9\linewidth]{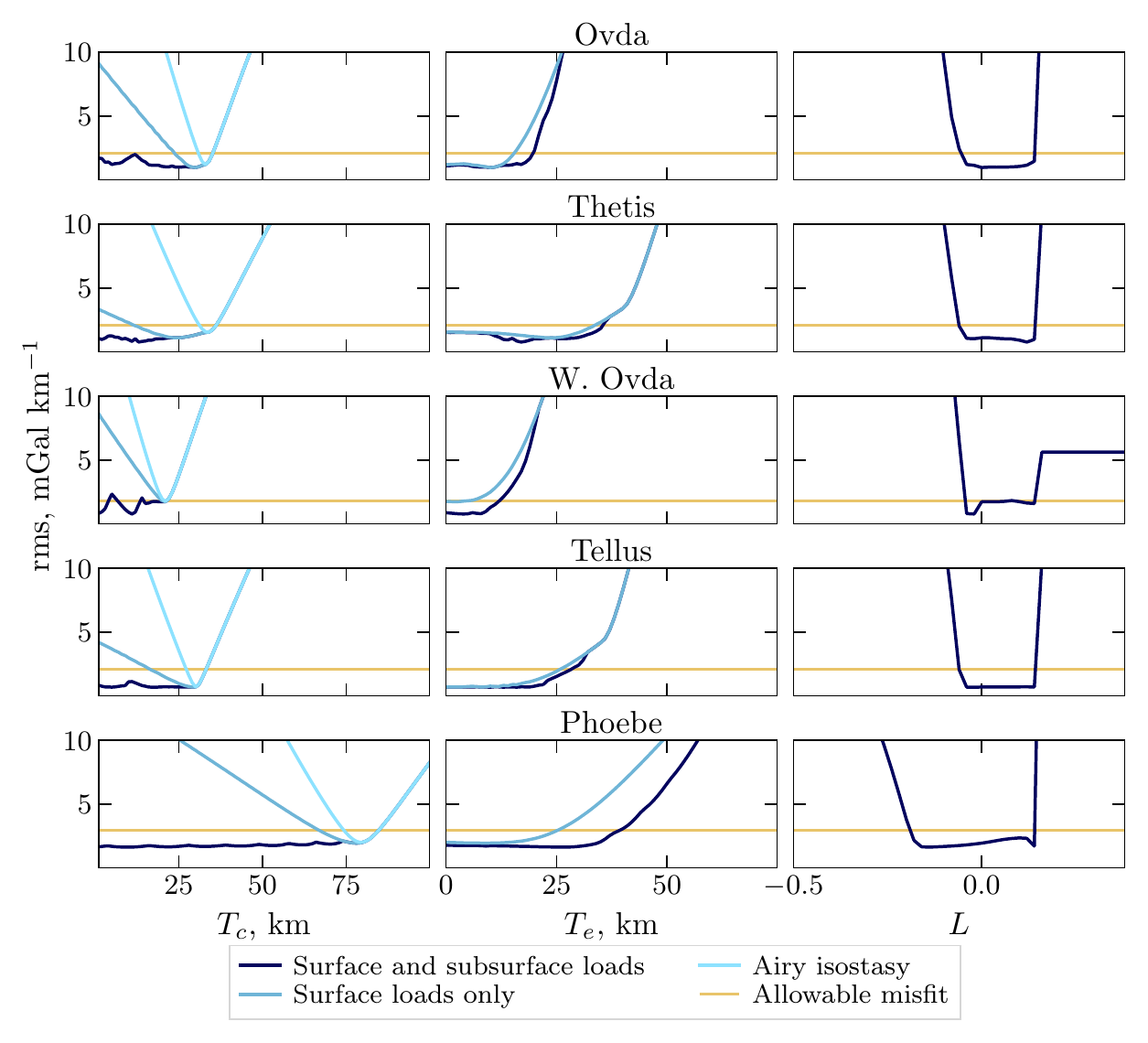}
	\caption{One-dimensional misfit plots for the crustal thickness (left), elastic thickness (center) and load ratio (right) for (from top to bottom) Ovda, Thetis, W. Ovda, Tellus, and Phoebe. Dark blue curves correspond to the case where both surface and subsurface loads are considered, light blue curves correspond to the scenario where only surface loads are included and the cyan curves correspond to the case of Airy isostasy.}
	\label{fig:misfit_allreg}
\end{figure}

In the following subsections, we present the crustal thickness, elastic thickness and load ratio estimations for the five regions. Similar to our analysis of Alpha Regio, these results consider three loading scenarios: both surface and subsurface loads, surface loads only, and a purely Airy isostasy regime. The misfit curves for each plateau are presented in Figure \ref{fig:misfit_allreg} and the estimated parameters are summarized in Table \ref{tab:results}.

\paragraph*{Thetis and Ovda Regiones.}

Thetis, Ovda and Western Ovda are adjacent to each other, and comprise the western part of Aphrodite Terra. Thetis and Ovda are the two highest crustal plateaus, with the latter reaching over 4 km altitude with respect to the surrounding plains. For these two regions the simple Airy isostasy model yields a satisfactory fit to the data, resulting in crustal thickness estimations of $34(32,36)$\,km and $33(31,35)$\,km, respectively  (values in parentheses here represent the +-1-sigma uncertainties). If we consider elastic support of surface loads only, the 1-sigma upper limit of the crustal thickness does not change, but the lower limit decreases, allowing a larger range of values than the simple Airy case. For this case, the crustal thickness of Ovda is $T_c = 30(24,35)$\,km and $T_c = 24(12,36)$\,km at Thetis. Similar to what was found for Alpha Regio, when subsurface loads are included it is only possible to obtain an upper bound for the crustal thickness. (In fact, this is the case for all analyzed regions.) For this scenario, the crustal thickness of Ovda is constrained to be less than 35\,km with a best-fit of $T_c=30$\,km while for Thetis the crustal thickness is constrained to be less than 36\,km with a best-fit of $T_c=13$\,km.

Regarding the elastic thickness, the scenarios of surface loads only and both surface and subsurface loads provide similar results. These two loading scenarios only allow for an upper bound of this parameter. In the case of Thetis we find that $T_e < 33$ km for the surface loading model and $T_e < 35$ km when subsurface loads are allowed, whereas for Ovda the upper limits are considerably smaller. In this region the top-loading scenario results in $T_e < 15$ km and the inclusion of subsurface loads results in $T_e < 19$ km. The range of accepted load ratio values is very similar for both regions, where $L =0.00( -0.04,0.14)$ for Ovda and $L=0.12(-0.06,0.14)$ for Thetis. Although the best-fitting load ratio is somewhat different for the two regions, in Figure \ref{fig:misfit_allreg} we can see that the range of acceptable values is very similar.

\paragraph*{Western Ovda Regio.}

Of the five regions discusses in this section, Western Ovda is the region that presented most difficulties to fit the Airy isostasy and surface-loading only models. In Figure \ref{fig:misfit_allreg} we see that in these two cases the minimum misfit values are very close to the threshold limit with both resulting in a crustal thickness of $T_c = 21$\,km. When subsurface loads are allowed, we can only define an upper limit for the crustal thicknesses of 21\,km. As for the elastic thickness, with the top-loading model we obtain $T_e < 4$\,km, and the addition of subsurface loads increases the upper limit to $12$\,km. In terms of the load ratio, we estimated $L =-0.02(-0.04,0.14)$ where we see that small amounts of buoyant material in the mantle ($-0.04<L<-0.02$) are clearly associated with the best fitting models, which is unusual among the plateaus in this study. Nevertheless, dense crustal intrusions with loading ratios greater than zero are also possible over a limited range.

It has been suggested that Western Ovda Regio is in fact not a typical crustal plateau, and that it instead represents a transition between the plateaus and tessera inliers \cite[e.g.][]{Nunes2004}. Its topography is characterized by an $\sim 2$\,km high semicircular rim with a low relief interior that  is largely embayed by volcanic plains. In particular, a 200\,km diameter corona is located in the volcanic plains of its interior. In addition,  W. Ovda has the lowest gravity signal among all crustal plateaus, and even has a negative anomaly in the center. The fact that our model is parametrized in terms of the topography and W. Ovda is fully collapsed at the center could explain why our surface-loading model does not work so well in this region. This is exacerbated by the fact that the localization window we use has higher amplitudes in the center where the low elevations are located. We note that in our analysis of Ovda and W. Ovda regiones we made use of two separated degree ranges to perform the inversion. We chose to invert for both ranges simultaneously since in the case where the two ranges were investigated separately we obtained consistent results but with uncertainties that were up to two times larger.

\paragraph*{Tellus Regio.}

Tellus Regio is an isolated crustal plateau located in the north hemisphere of Venus. The inversion results at Tellus are very similar to the estimations for Ovda and Thetis. When assuming Airy isostasy we obtain crustal thickness estimations of $T_c = 30(28,33)$\,km. When including only surface loads the 1-sigma upper limit remains unchanged but the lower limit decreases from 28\,km to 17\,km. Finally, in the case where subsurface loads are allowed we were only able to obtain an upper bound on the crustal thickness of 33\,km. The elastic thickness is constrained to be less than 25 km when only surface loads are considered, and less than 28\,km when both surface and subsurface loads are considered.  Similar to the previous analyses, load ratios ranging from $-0.06$ to $0.14$ are accepted.

\paragraph*{Phoebe Regio.}

Phoebe Regio presents the most distinct results in comparison to the other studied plateaus. Although the region has been previously defined as a crustal plateau \cite[e.g.][]{PhillipsHansen1994, Nunes2004} it presents some important differences, and has been characterized as a hybrid between plateaus and volcanic rises \cite[e.g.][]{Phillips1998, Kiefer2003}. In terms of tectonic features, Phoebe presents a unique structural fabric in which ribbon terrains, typical in other plateaus, are not present \citep{Phillips1998}. In addition, Phoebe is connected to the volcanic rise Beta Regio by an extensive rift system called Devana Chasma. Previous geophysical studies \cite[e.g.,][]{simons_1997, Kiefer2003} have also shown that the gravity signal at Phoebe is somewhat in-between volcanic rises and crustal plateaus, probably being partially supported by a mantle plume. Of the regions in this analysis, Phoebe contains that highest values of the admittance, with values increasing from 30\,mGal\,km$^{-1}$ at low degrees to 60\,mGal\,km$^{-1}$ at high degrees. The highest admittances are about two times larger than those of the other plateaus in our study.

The three investigated scenarios provide acceptable fits for Phoebe Regio. When Airy isostasy is considered the crustal thickness is $79(75, 84)$ km, whereas when surface loads are considered, the range of values increases slightly to $78(68, 84)$ km. We note that these values are about two to three times higher than the values found in other plateaus. Nevertheless, when subsurface loads are considered, all values for the crustal thickness less than 84 km are allowed. The elastic thickness is found to be less than 25 km when surface loads are considered, and less than 39 km with both surface and subsurface loads are considered. As for the loading ratio $L$, we note that Phoebe allows for considerable subsurface loads. Whereas the other regions in this study possess negative values down to $-0.06$, for Phoebe, this increases downward to $-0.18$. Though we can not distinguish between surface and subsurface loads, our study corroborates the interpretation that Phoebe is potentially associated with a much more important mantle plume or buoyant layer than the other crustal plateaus.

\begin{table}[h]
	\caption{Summary of results for the six studied plateaus. $L \neq 0$ corresponds to the scenario where surface and subsurface loads are allowed, $L=0$ has only surface loads and $T_e=0$ represents the Airy isostasy case. The values in parentheses indicate the 1$\sigma$ limits. In cases where the limits are not present, no acceptable fits were found, with the value shown corresponding to the best-fitting value.}
	\centering
  \begin{tabular}{lcccccc}
    \hline
    \multirow{2}{*}{Region} &
      \multicolumn{3}{c}{$T_c$ (km)} &
      \multicolumn{2}{c}{$T_e$ (km)} &
      \multicolumn{1}{c}{$L$} \\
		\cmidrule(lr){2-4} \cmidrule(lr){5-6} \cmidrule(lr){7-7}
    & $L \neq 0$ & $L=0$ & $T_e=0$ & $L \neq 0$ &  $L=0$ & $L \neq 0$\\
    \hline
		Ovda& 30(1--35)& 30(24--35)& 33(31--35)& 10(0--19)& 10(0--15)& 0.00(-0.04--0.14) \\ [5pt]
		Thetis& 13(1--36)& 24(12--36)& 34(32--36)& 17(0--35)& 24(0--33)& 0.12(-0.06--0.14) \\ [5pt]
		W. Ovda& 11(1--21)& 21(20--22)& 21& 4(0--12)& 2(0--4)& -0.02(-0.04--0.14) \\ [5pt]
		Alpha& 15(1--21)& 15(9--21)& 24& 20(9--24)& 20(12--24)& 0.00(-0.02--0.14) \\ [5pt]
		Tellus& 18(1--33)& 30(17--33)& 30(28--33)& 10(0--28)& 2(0--25)& -0.02(-0.06--0.14) \\ [5pt]
		Phoebe& 10(1--84)& 78(68--84)& 79(75--84)& 26(0--39)& 9(0--25)& -0.14(-0.18--0.14) \\ [2pt]
    \hline
  \end{tabular}
	\label{tab:results}
\end{table}

\section{Discussion} \label{sec:discussion}

\subsection{The Crustal Thickness of the Highland Plateaus}\label{sec:disctc}

Several studies have used gravity and topography data to investigate the Venusian crustal plateaus. Most of those, notably in the Magellan Era, studied the crustal thickness of these features making use of spatial techniques and assuming that they were isostatically compensated \citep{smrekar1991, kuncinskas_1994, moore_1997}. The crustal thickness of the highland plateaus was also investigated in some of the earliest developments of localized spectral admittance analyses \citep{grimm_1994, simons_1997}. \cite{anderson_2006} created spectral classes from localized spectral admittances to estimate the crustal thickness and elastic thickness across the planet. \cite{james_2013} developed a global compensation model separating the effects from shallow compensation, related to crustal thickness variations, and dynamic compensation mechanisms associated with mass anomalies in the mantle. From this model the authors were able to construct a global crustal thickness map assuming a mean crustal thickness of 15 km. Finally, \cite{jimenezdiaz_2015}  made use of the crustal thickness modeling introduced by \cite{Wieczorek1998} to generate a crustal thickness map of Venus assuming an average crustal thickness of 25 km. In contrast, our study made use of three loading scenarios, with different levels of complexity, to investigate the crustal and lithospheric structure and the possible compensation mechanisms of the crustal plateaus.

Our investigation showed that using an Airy isostasy regime to study the crustal plateaus, which has been done in many previous works, is in most cases a valid approximation to fit the admittance within uncertainties, with the exception of Alpha Regio where some flexural rigidity is necessary to properly fit the data. We added a level of complexity to our model by assuming that surface loads are supported elastically by the lithosphere, i.e. the elastic thickness can be different from zero. From this, we were able to confirm that the elastic thicknesses associated with the crustal plateaus are low, with a best-fitting average of approximately 15 km among all regions. Considering the uncertainties, we found that most regions are consistent with $T_e=0$ km with the exception of Alpha Regio where we obtained a lower bound of $T_e=12$ km. Moreover, we found upper bound values of no more than 35 km.

The final level of complexity in our study was to assume that the surface relief is a combination of surface loads and support from either a buoyant mantle layer or a high density intrusion within the crust. We found that the best-fitting load ratio is equal to zero or has small positive values for most regions, indicating a possibility for small amounts of dense intrusions in the crust. Considering the uncertainties, a small negative loading ratio up to about $-0.05$ is permitted for Ovda, Thetis and W. Ovda and Tellus regiones. The only exception is Phoebe Regio where $L$ can go as low as $-0.18$ with a best-fitting value of $-0.14$. To obtain a more physically meaningful estimation we can convert $L$ to the ratio of subsurface to surface loads, $f$ (see eq. \ref{eq:f}). We find that Phoebe may have a buoyant mantle load that is about 12\% (up to 15\%) of the surface load while in other regions the fraction is constrained to a maximum of 4\%. Overall, our results show that the interior structure of the plateaus is well-described by the inclusion of surface loads only with a small, arguably negligible, contribution of flexural support, which is consistent with the interpretation that the geological processes responsible for their formation are no longer taking place.

\begin{figure}[h!]
  \centering
	\includegraphics[width=\linewidth]{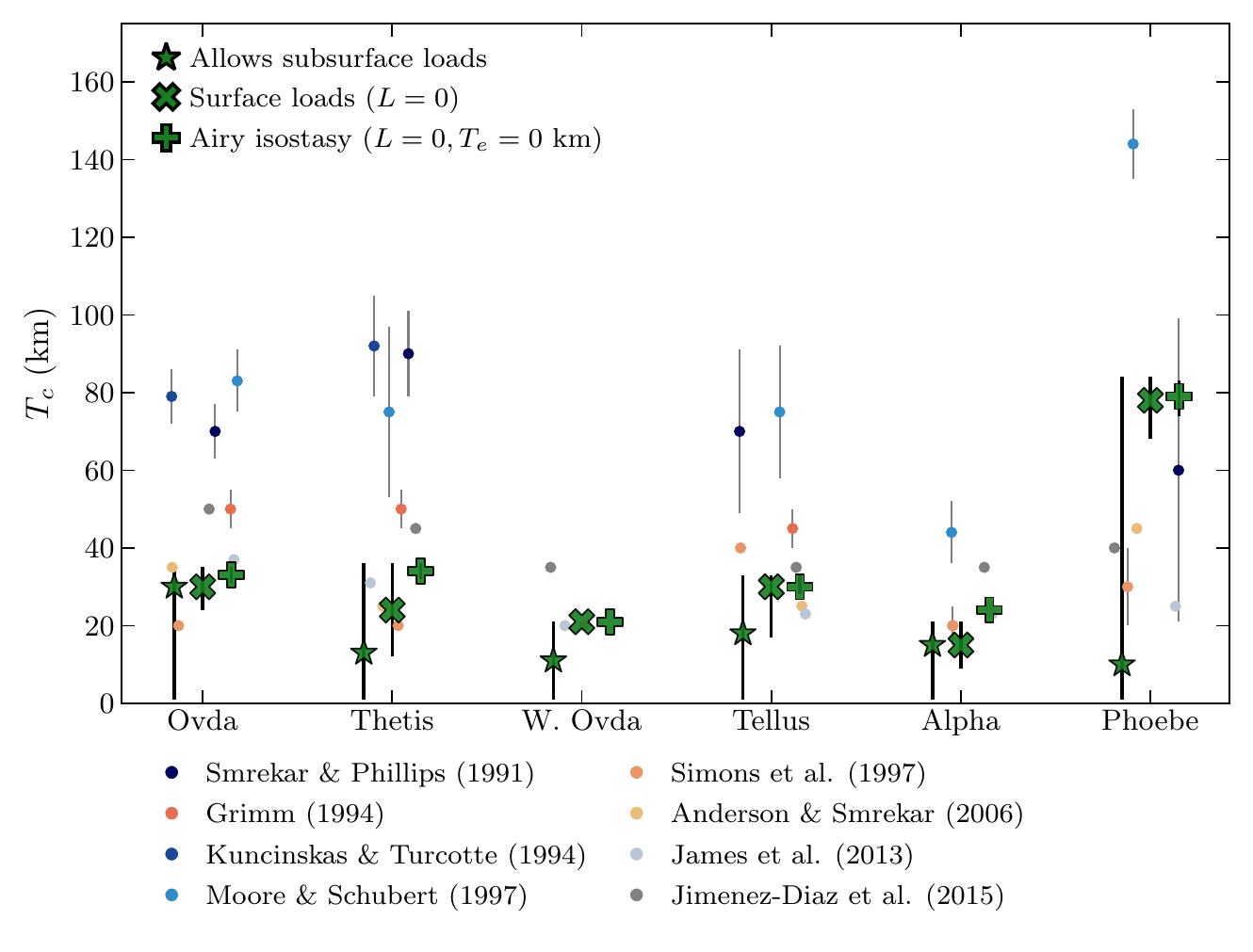}
	\caption{Crustal thickness estimates for the crustal plateaus of Venus. The values in green represent our results considering the three investigated compensation scenarios: Airy isostasy ($+$), surface loading ($\times$), and surface-subsurface loading ($\star$). The dots correspond to estimates from previous studies. Dark and light blue are associated with the use of spatial analyses methods, dark and light orange correspond to localized spectral analyses and greys represent global crustal thickness modeling.}
	\label{fig:compare_results}
\end{figure}

Considering the three loading scenarios just described, Figure \ref{fig:compare_results} presents our crustal thickness estimations for the six studied plateaus. The figure also includes crustal thickness estimates from the previous studies mentioned above. Looking at our crustal thickness estimations we can see that for Ovda, Thetis, W. Ovda, Tellus and Alpha the results obtained are quite similar, with best-fitting values ranging from 15 to 30 km in the surface-loading scenario and from 11 to 30 km when internal loads are included. In addition, in these regions the different loading scenarios do not have a major impact on the results. Overall, the most important difference between the three loading scenarios is the accepted range of values. As one would expect, decreasing the number of free parameters reduces the uncertainties of the estimations. Assuming an Airy isostasy regime results in good fits for most regions with very low uncertainties of roughly $\pm 2$ km. When only surface loads are allowed we found acceptable crustal thickness values for every region and all estimations include a lower bound. For this case, the crustal thicknesses were found to be uncertain by about $\pm 6$ km. Lastly, when including both surface and subsurface loads, we note that we were able to obtain only an upper bound on the crustal thickness. When comparing the three scenarios, we note that the crustal thicknesses obtained when assuming Airy isostasy can be slightly larger than that of the other two techniques, with best-fitting values  up to 10 km larger with respect to the surface-loading scenario.

Phoebe Regio is the only region where our crustal thickness estimates are different than the other crustal plateaus. When internal loads are not included the admitted crustal thickness values  range from 68 to 84 km, which corresponds to roughly three times the thickness found for the other regions. However, when subsurface loads are present the uncertainties are increased dramatically. Even though the upper bound in this case is comparable to the other two models, we do not find a firm lower limit for the crustal thickness. The best fitting value is 10 km, which is about 70 km lower than for the Airy and surface loading models, is more consistent with the crustal thicknesses obtained for the other plateaus.

Comparing our estimations with the previous studies we find that, in general, the more recent investigations present a good agreement with the values we have obtained. On the other hand, many investigations done in the 1990s present considerably higher values, ranging from twice to three times what we find. This discrepancy between the earliests and latest studies probably arises from the combination of several factors, but we expect that the main contributor for the observed differences are related to differences in methodology. Many early studies assumed that the plateaus were isostatically compensated, which can bias the crustal thickness estimations towards larger values. However, even when comparing our Airy isostasy estimation with previous ones an important discrepancy persists, particularly for the studies that made use of spatial techniques. Therefore, it is likely that the use of different analysis techniques has a major impact on the estimations. With our approach we were able to select only the portion of the spectrum where the correlation between gravity and topography is high. On the other hand, in space domain approaches, it is not possible to do this selection  since this type of analysis collapses the wavelength-dependent gravity-topography ratio into a single value. The inclusion of gravity signals from uncorrelated sources or from long-wavelength mantle signals could be part of the cause of the bias of their crustal thickness estimations. It is also important to remark that our study presents important improvements in comparison to most previous studies regarding the uncertainty analysis. In fact, several prior studies did not provide any uncertainties for their estimations.

A second possible source for the observed discrepancy is the use of different datasets.  The early gravity studies relied on gravity models with substantially lower resolutions than the MGNP180U model used in this work (which was published in 1999). Particularly, \cite{smrekar1991} only had access to a gravity model from the Pioneer Venus mission, while the other studies done in the 1990s used a variety of preliminary gravity models with maximum resolutions ranging from spherical harmonic degree 60 to 120, depending on the publication year. A historical review about these early gravity models of Venus can be found in \cite{Sjogren1997}.

\begin{figure}[h!]
  \centering
	\includegraphics[width=\linewidth]{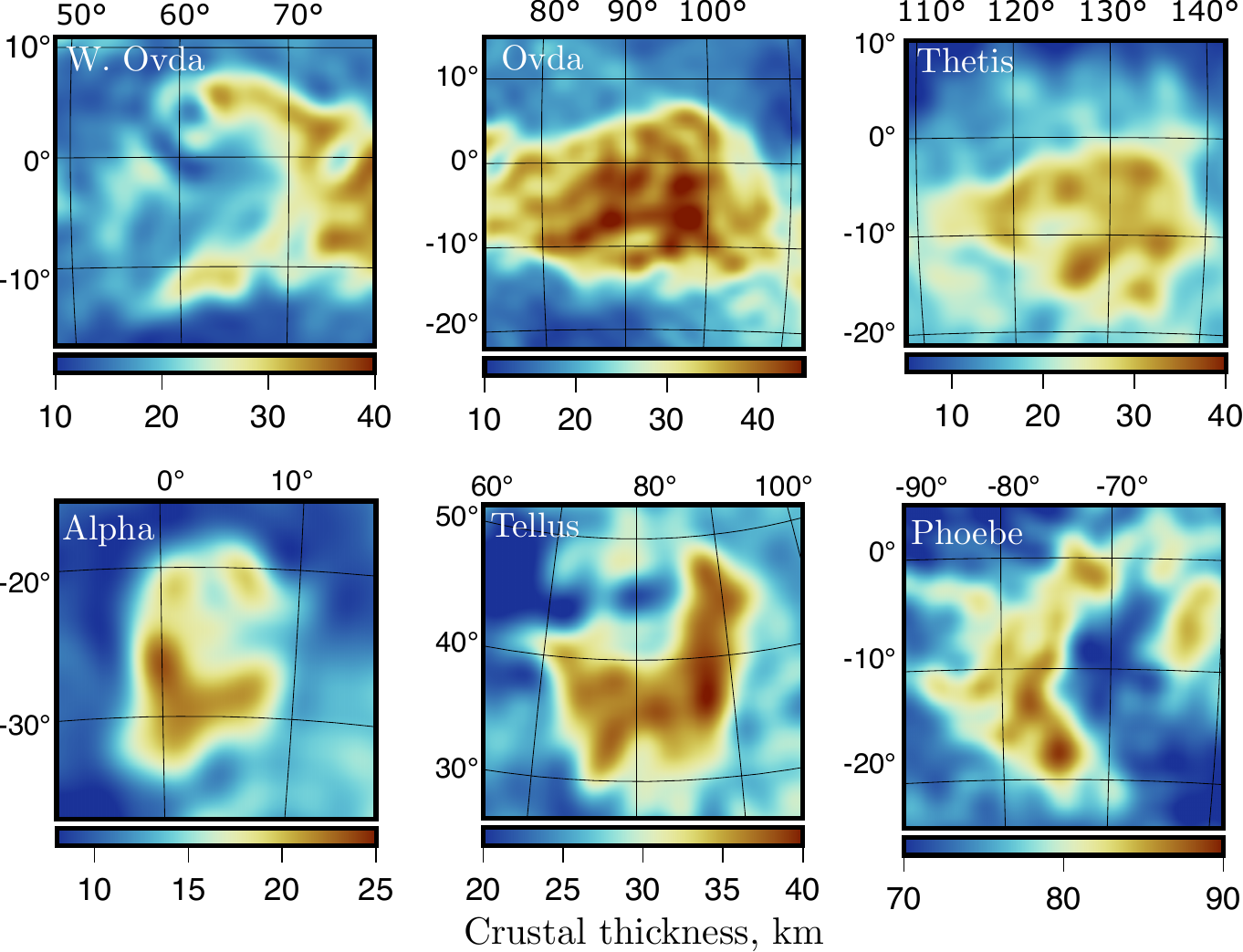}
	\caption{Predicted local crustal thickness variations of the six studied crustal plateaus. These maps make use of the surface topography, the best fitting crustal thickness, and the predicted flexure of the crust-mantle interface for the best-fitting elastic thickness at each region for the surface-loading model. The adopted projection for these maps is Lambert azimuthal equal-area. }
	\label{fig:localcrusthick}
\end{figure}

Up until this point we have reported crustal thickness estimations that correspond to the average crustal thickness of each region (weighted by the amplitude of the localization window). However, it is also interesting to investigate the local variations of crustal thickness within each region. In order to do so, we employ the best-fitting elastic thickness and crustal thickness estimates obtained using the surface-loading scenario (Table \ref{tab:results}). With these values, we then estimate the lateral variations in deflection of the lithosphere (i.e., $w$ in Figure 3). From this global map of the crust-mantle interface, we then estimate the average crustal thickness of the analysis region (weighted by the localization window) and compare with the value $T_c$ that we obtained from our admittance analysis. Finally, we modify the average depth of the interface $w$ (i.e., the degree-0 spherical harmonic term) such that the predicted value is the same value as from our analysis. The resulting crustal thickness maps, based on the best-fitting crustal thickness and elastic thickness values, are shown in Figure \ref{fig:localcrusthick}. These maps show that, due to the crustal roots, crustal material can reach depths much larger than the regional average. Particularly at Ovda, the highest plateau, the crustal thickness can reach up to 54 km considering the uncertainties of the analysis parameters.

We emphasize two aspects concerning our crustal thickness models presented in Figure \ref{fig:localcrusthick}. First, even though these are computed using spherical harmonic coefficients of the deflection of the crust-mantle interface, the model is only valid locally within the analysis region. Second, our analysis differs from the more common technique where the relief of the crustal mantle interface is inverted in order to satisfy the observed gravity field \cite[e.g.][]{Wieczorek1998, james_2013, jimenezdiaz_2015}. In particular, in our model, the relief along the crust mantle interface is predicted from a flexure model that satisfies the admittance over a limited range of spherical harmonic degrees.

From the regional average crustal thickness at each plateaus $T_c$ we are also able to estimate the crustal thicknesses at the mean planetary elevation, $T_0$. Given that the crustal plateaus are overall consistent with being in Airy isotasy,  we can use the following relation:
\begin{equation}
		T_c = T_0 + \overline{h}_w \left( 1 + \frac{\rho_c}{\rho_m - \rho_c} \right)
\end{equation}
where $\overline{h}_w$ is the window-weighted average elevation of the analysis region with respect to the mean planetary radius. Disregarding the anomalous $T_c$ estimation of Phoebe Regio, we find that the average thickness of the crust, $T_0$, is 22 km on average. Of course, this approach assumes that the crustal density is constant. In Table S2 we provide, for all studied plateaus, the regional average topography weighted by the localization window, the implied mean crustal thicknesses $T_0$ and the maximum depths reached by crustal materials. We note that $T_0$ can also be derived surface-loading scenario from the degree-0 term of each crustal thickness map (Figure \ref{fig:localcrusthick}). In this case, for the best-fitting crustal thickneses and elastic thickness values, the average $T_0$ is 17 km.

\subsection{Heat Flow at the Crustal Plateaus} \label{sec:disheatflow}

The elastic thickness is strongly related to the thermal state of the lithosphere and, in the absence of direct heat flow measurements, it is one of the few quantities that can help constrain the thermal evolution of the planet. It is important to remark that these estimations are associated with the lithospheric thermal properties during the formation of plateaus and do not necessarily correspond to their current, probably colder, thermal state \citep{Albert2000}. The estimation of the heat flow from the elastic thickness is based on the method introduced by \cite{McNutt1984}, which equates the bending moment of an idealized elastic plate with the same bending stresses within a more realistic rheology model that includes yielding, commonly known as the mechanical lithosphere. In the elastic plate model, stresses vary linearly with depth and the bending moment depends on the curvature $K$ and flexural rigidity of the plate. On the other hand, for the mechanical lithosphere, which assumes a elastic-plastic model, the stresses are governed by brittle failure \cite[e.g.][]{Mueller1995} in the upper and colder lithosphere and viscous stresses \cite[e.g.][]{McNutt1984} in the lower and hotter lithosphere.

In practice, in order to estimate the bending moment of the mechanical plate that includes yielding, the numerical integration is limited to a depth where the stress difference becomes negligible and the lithosphere loses its mechanical strength. We consider a bounding stress value of 50 MPa. This choice is somewhat arbitrary but for values smaller than 100 MPa the impact is small since the viscous stress law has an exponential dependency on the temperature \citep{McNutt1984}. The dependency of the viscous flow on the temperature is, in fact, crucial for the determination of the heat flow from the yield stress envelope. Assuming that the temperature gradient $dT/dz$ across the lithosphere is constant we are able to vary this parameter until the best match between the two estimates of the bending moment. Then, we can estimate the heat flow $F$ making use of the relation $F = k \, dT/dz$, where $k$ is the assumed thermal conductivity of the lithosphere. Moreover, we can compute the thickness of the mechanical lithosphere as the depth where the viscous stress reaches the defined bounding stress of 50 MPa associated with the best-matching yield stress envelope.

The stress laws adopted also depend on the strain rate and the mineralogical composition of the lithosphere. As discussed in \cite{Brown1997}, during the formation of tessera terrains craters were being destroyed faster than they were forming which indicates a high strain rate, likely around $10^{-15}$ s$^{-1}$ which is the value picked for our investigation. Moreover, in order to compute the elastic bending moment we adopt the maximum curvature found at each region for each set of model parameters as advocated by \cite{Mueller1995}. The maximum curvature is computed by taking the second derivative of the modeled lithospheric deflection $w$. We remark that the curvature values are derived from flexural models and not actual observations. Hence, the curvature depends upon the validity of the model assumptions.

We adopt a thermal conductivity of 2 W m$^{-1}$ K$^{-1}$ which is a typical value for terrestrial basalts \cite[e.g.,][]{Clifford1985, Clauser1995}. We note, however, that the thermal conductivity of rocks strongly depends on many parameters such as porosity, fluid content, composition, and temperature. Previous studies that investigated the lithospheric heat flow on Venus considered a large variety of thermal conductivity values, ranging form 2 W m$^{-1}$ K$^{-1}$ \citep{Bjonnes2021} to 4 W m$^{-1}$ K$^{-1}$ \citep{ORourke2015} and it is important to take this into account when comparing results from different studies. Concerning the composition, we consider a dry diabase rheology for the crust and a dry olivine rheology for the mantle, making use of the same flow law parameters as in \cite{Resor2021}. Since new studies suggest a felsic rheology for the crustal plateaus \citep{Gilmore2015}, we also performed heat flow estimations considering an anorthite (plagioclase) rheology for the crust. We found that this change in rheology had a minor impact on the heat flow estimations, corresponding to a maximum increase of 3 mW m$^{-2}$ with respect to the diabase rheology.

We estimate the best-fit heat flow associated with the best-fit elastic thickness for the six plateaus, and the associated uncertainties are computed using the 1-sigma uncertainties of the elastic thickness. Given that $T_e = 0$ km corresponds to an infinite thermal gradient, we were unable to compute an upper bound of the heat flow for most regions. In Table \ref{tab:heatflow} we summarize the curvature, mechanical thickness, thermal gradient and heat flow estimations that we obtained. As expected, the mechanical thickness estimations are always larger than the associated elastic thicknesses, ranging from 1.5 to 2 times their values. The best-fitting thermal gradients and heat flows vary considerably among the plateaus, ranging from 8 to 100 $^{\circ}$C\,km$^{-1}$ and 16 to 200 mW m$^{-2}$ . This difference is less striking when we look into the lower limits, where the thermal gradient varies from 7 to 10 $^{\circ}$C\,km$^{-1}$ and the heat flow from 12 to 20 mW m$^{-2}$. The only exception is for W. Ovda Regio that has lower bounds of 45 $^{\circ}$C\,km$^{-1}$ for the temperature gradient and 90 mW m$^{-2}$ for the heat flow.

\begin{table}[h]
	\caption{Heat flow and related quantities for the crustal plateaus of Venus. $K$ is the maximum plate curvature, $T_m$ is the mechanical lithosphere thickness, $dT/dz$  is the temperature gradient, and $F$ is the surface heat flow. The values in parentheses represent the 1$\sigma$ limits of each estimation.}
	\centering
  \begin{tabular}{lcccccc}
    \hline
    Region & $K$ ($10^{-7}$ m$^{-1}$)  & $T_m$ (km) & $dT/dz$ ($^{\circ}$C\,km$^{-1}$)&  $F$ (mW m$^{-2}$)\\
    \hline
		Ovda& 5.35(4.99--5.56)& 16(0--30)& 16(11--)&31(21--) \\ [5pt]
		Thetis& 3.9(3.19--5.23)& 39(0--48)& 8(7--)&16(13--) \\ [5pt]
		W. Ovda& 5.2(5.19--5.21)& 3(0--6)& 92(45--)&184(89--) \\ [5pt]
		Alpha& 2.2(2.04--2.45)&27(18--32)&12(10--19)&24(20--37) \\ [5pt]
		Tellus& 3.08(2.23--3.08)& 3(0--35)& 102(10--)&203(18--) \\ [5pt]
		Phoebe& 3.75(2.94--3.83)& 13(0--38)& 20(7--)&39(13--) \\ [2pt]
    \hline
  \end{tabular}
	\label{tab:heatflow}
\end{table}

Overall, our heat flow estimations have very large uncertainties, mostly being constrained to be only larger than about $ 15$ mW m$^{-2}$. Therefore, the interpretability of these results is quite limited. Alpha Regio, however, is an exception because it is the only region where we found a lower bound for the elastic thickness that constrains the heat flow within the range 20--37 mW m$^{-2}$. For the purposes of discussion, we will assume that this range is representative of the other highland plateaus, and then compare these estimations to those obtained using independent techniques In order to perform a consistent comparison, we rescaled all heat flow estimations to use the same thermal conductivity as in this study.

Global thermal evolution models and previous elastic thickness investigations suggest that the current average or ``ambient'' heat flow on Venus is around 10--30 mW m$^{-2}$ \citep{Solomatov1996, Phillips1997, ORourke2015}. Furthermore, hydrocode modeling of the Mead impact basin formation from \cite{Bjonnes2021} constrained the temperature gradient to be 6 to 14 $^{\circ}$C\,km$^{-1}$, corresponding to a heat flow of 12--28 mW m$^{-2}$. This latter result was proposed to be independent of location and representative of the past 300 Myr to 1 Gyr. Gravity and topography studies of Venusian volcanic rises, including Atla, Bell and Eistla regiones, lead to heat flows estimations of 21 to 35 mW m$^{-2}$ \citep{Phillips1997}, which are very similar to our results for Alpha Regio.  In addition, flexural studies based on topography data indicate that coronae are associated with major heat flow anomalies, reaching up to $\sim 100$ mW m$^{-2}$ \citep{ORourke2018, Russell2021}, while steep-sided domes are associated with regional heat flows of 40 mW m$^{-2}$ \citep{Borrelli2021}.

Our heat flow estimates generally overlap with the predicted present-day global average values.  Nevertheless, our estimates are on the high end of these predictions, and are more similar to those obtained for the volcanic rises. It is possible that the plateaus were associated with higher heat flows at the time of their formation.  At least for Alpha Regio, we find that the excess heat flow with respect to the estimated global average is about 10 mW m$^{-2}$. For comparison, this 10 mW m$^{-2}$ excess heat flow is consistent with estimations of excess heat flow associated with hotspots on slow moving plates on Earth \cite[e.g.,][]{Sleep1990}. Furthermore, orogenic belts on Earth's continental crust have also been associated with surprisingly low elastic thicknesses (lower than 20 km) and high heat flows \cite[e.g.][]{McNutt1988, Burov1990} which compares favorably with the values we found for the crustal plateaus. The cause of these anomalously low elastic thicknesses in continental regions is not fully understood, but data inversion and finite-element modeling studies indicate that decoupling of the strong upper crust and upper mantle probably is an important contributor \citep{McNutt1988, Burov1995, Brown2000}. In fact, we also observed crust-mantle decoupling in the estimated yield-stress envelopes of some crustal plateaus in a few cases. The possibility of crust-mantle decoupling on Venus has been previously discussed in \cite{Buck1992}, \cite{Azuma2014}, and \cite{Ghail2015}.

Regarding previous heat flow estimations of crustal plateaus, \cite{Brown1997} and \cite{Resor2021} were able to constrain the heat flow associated with the formation of folds in tessera terrains based on geodynamic modeling of tessera deformation. In short, they estimated the depth of the brittle-ductile transition based on the dominant wavelength of regularly spaced contractional ridges and the crustal rheology. These investigations showed that the folds observed in these regions are associated with thermal gradients of roughly 20--25 $^{\circ}$C\,km$^{-1}$, corresponding to heat flows of 40--50 mW m$^{-1}$, for a diabase rheology. For a felsic rheology, \cite{Resor2021} estimated that thermal gradients would be approximately twice as high, with heat flows ranging from 90 to 100 mW m$^{-1}$. These estimated heat flows are considerable higher than what we obtained for Alpha Regio, but are nevertheless compatible with the uncertainties of the other crustal plateaus.

\subsection{Insights on the Thermal and Geological Evolution of the Crustal Plateaus}\label{sec:discevol}

\begin{figure}[h!]
  \centering
	\noindent\includegraphics[width=\linewidth]{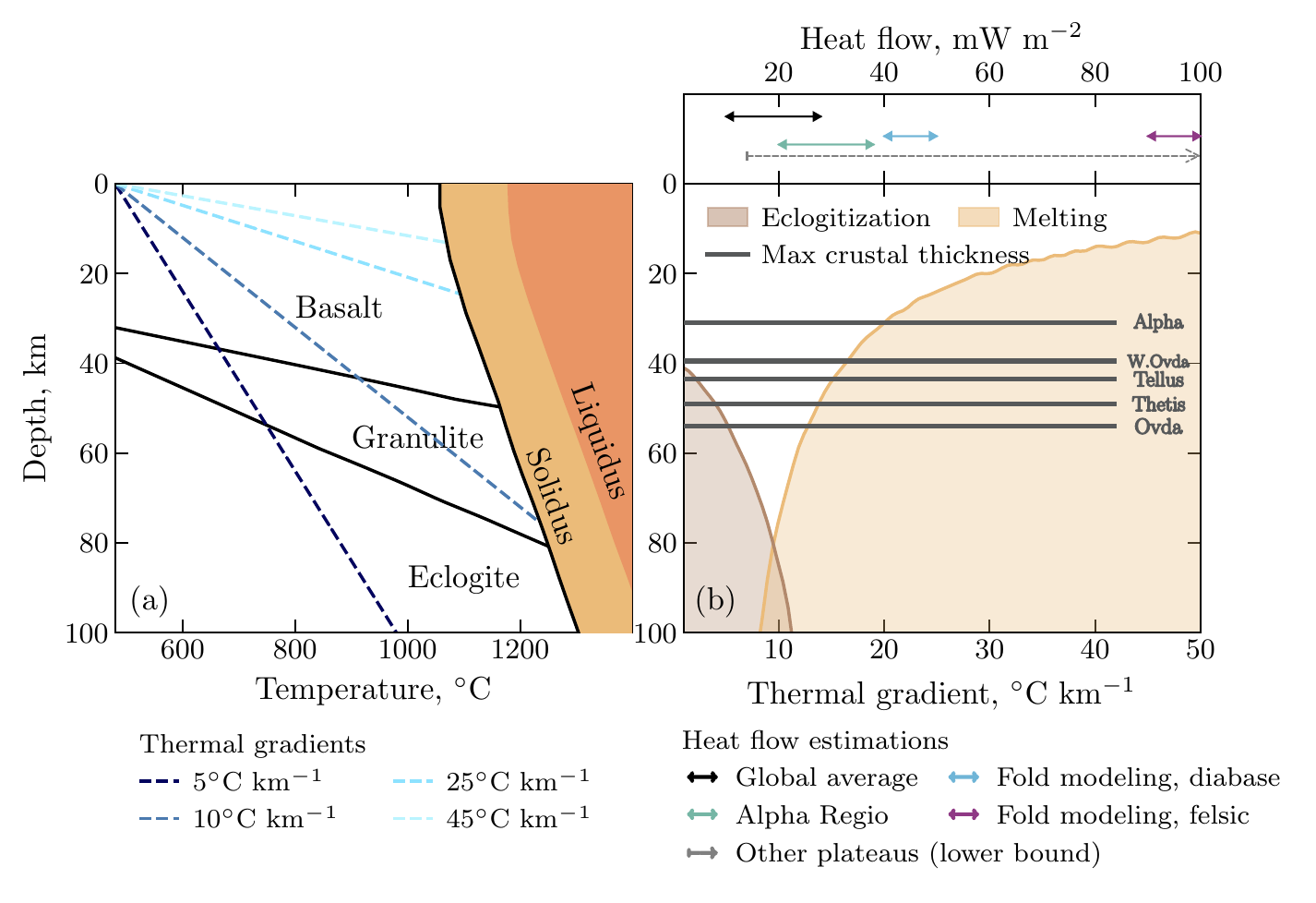}
	\caption{(a) Phase transitions in the basalt system as a function of depth and temperature. The phase transitions, solidus and liquidus are adapted from \cite{Hess1990} and dashed curves represent temperature profiles with constant temperature gradients from 5 to 45 $^{\circ}$C/km. (b) Depth range of eclogitization (brown) and melting (orange) as a function of the thermal gradient and corresponding heat flow. The horizontal lines correspond to the maximum depths reached by crustal materials at each plateau given the $1\sigma$ limits of average crustal thickness and elastic thickness. These maximum depths consider lithospheric flexure and are hence greater than the average values reported in the main text. The arrows above the plot correspond to different heat flow estimations associated with the crustal plateaus. The estimate we obtained for Alpha Regio is shown in green, whereas the gray arrow corresponds to the average lower bound for the other plateaus. The blue and purple arrows represent estimations from geodynamical modeling of fold formation for a diabase rheology \citep{Brown1997, Resor2021} and felsic rheology \citep{Resor2021}, respectively. The black arrow corresponds to global average estimations.}
	\label{fig:phasetransition}
\end{figure}

Crust-constituent minerals may undergo solid-state phase transitions or melt when they reach certain pressure-temperature conditions. These phase transitions affect the density of the host materials, and may affect the dynamics of the crust and underlying mantle. In Figure \ref{fig:phasetransition} we plot the conditions where the main phase transitions associated with a basalt system take place. Plot (a) shows the depth-temperature relationship for basalt phase transitions to granulite, eclogite, and the onset of melting. In (b) we plot the depths at which the eclogite phase transition and melting occur for as a function of the thermal gradient, which is assume to be linear across the lithosphere

For low thermal gradients ($\leq 10^{\circ}$C\,km$^{-1}$) basaltic compositions transitions into granulite at depths ranging from 30--40 km which in turn transition into eclogite when a depth of 50--80 km is reached. These transitions are mainly driven by the production of garnet from plagioclase present in the basaltic rocks. Granulite corresponds to a phase where only part of plagioclase has been transformed into garnet, i.e., both minerals are still on the system, while in the eclogite phase all plagioclase has transformed into garnet. Meanwhile, in an environment subject to higher thermal gradients, roughly above  10$^{\circ}$C\,km$^{-1}$ (or 20 mW m$^{-1}$), partial melting of the material occurs before the eclogite phase transition is reached. For reasonable depth ranges, the solidus corresponds to temperatures somewhere between 1000 to 1200$^{\circ}$C.

Combining the local crustal thickness estimations (Figure \ref{fig:localcrusthick}) with the heat flow estimations (section \ref{sec:disheatflow}), we are able to investigate whether basal melting of the crust or the formation of eclogite could have occurred at the base of the plateaus. These conditions are visualized in Figure \ref{fig:phasetransition}(b) where we show the maximum crustal thickness of each region and several heat flow estimations associated with these regions. Making use of model present-day global average estimations \cite[e.g.,][]{Solomatov1996,Phillips1997,ORourke2015, Bjonnes2021}  we adopt $\sim$10--28 mW m$^{-1}$ as a reasonable range of the current surface heat flow of Venus. Considering this range, represented by the black arrow, we find that the eclogite phase transition could be reached at the base of the two highest plateaus, Ovda and Thetis. In particular, we do not expect the eclogite phase transition to be reached in the other plateaus Tellus, W. Ovda, and Alpha regiones, nor for any of the plateaus if the heat flow there is greater than about 20\,mW\,m$^{-2}$.

One important property of eclogite is that its density ($\sim 3500$ kg m$^{-3}$) is considerably higher than basalt and likely higher than the underlying mantle, which would enable these materials to delaminate and sink into the mantle. Therefore, it is conceivable that delamination processes are potentially taking place in these regions, or that they occurred throughout their evolution. In fact, delamination could possibly explain the correlation and admittance decrease around spherical harmonic degrees 50--60 observed in several regions (see Figure \ref{fig:allreg_obs_win}). Moreover, as a result of delamination, the depth of the eclogite transition could potentially correspond to the maximum expected crustal thickness of the planet. Because of this, estimations of depths of the crust-mantle interface considerably greater than 70 km should be examined with suspicion, such as with the highest values (given the uncertainties) we obtain for Phoebe Regio.

The heat flows associated with the formation of the plateaus were likely higher than the present global average heat flow, as shown in Figure \ref{fig:phasetransition}(b). The green arrow indicates our heat flow estimation for Alpha Regio, which corresponds to the period of load emplacement in the region. The estimations based on fold modeling \citep{Brown1997, Resor2021} are shown in blue and purple for dry diabase and dry felsic rheologies, respectively. Though the present day globally averaged heat flow is probably not sufficient to cause melting in the crustal plateaus, the heat flows were probably substantially higher when the plateaus were forming (as obtained from our results and those of fold formation). Crustal materials that reached 40 km depth or more would have likely gone through some degree of melting. Considering the heat flow constraints from the formation of folds, the melting of crustal materials would happen at shallow depths, of about 25 km for diabase and 15 km for more felsic compositions. We note that the heat flow estimations based on fold modeling strongly depend on the rheology law used \citep{Brown1997}, appearing to be more sensitive to this parameter than the method by \cite{McNutt1984} used in our study. Hence, in the case of the fold modeling approach, we need better constraints on the the composition and volatile content in these regions in order to estimate the inferred heat flow. We also remark that these phase transitions are based on a basaltic system and more felsic systems would tend to melt at even lower temperatures than what is shown in Figure \ref{fig:phasetransition}. Nevertheless, these results already indicate that magmatic processes may have played an important role in the early formation of many crustal plateaus.

The substantial difference between heat flow estimations based on fold modeling and our estimation for Alpha Regio is somewhat puzzling. Nevertheless, even if we consider that the rheological flow laws used are correct, there are several factors that could help explain this discrepancy. Since the formation of folds is associated with higher heat flows, it is plausible that these features developed early in the plateau formation history when the crust was considerably thinner than the current observations. In fact, the very limited amount of volcanism in these plateaus could indicate that the crustal thickness was around 20 km or less at that time. We also remark that the heat flow estimations from fold formation are associated with a single fold wavelength that are not necessarily responsible for the entire thickening process of the plateaus.

It is also possible that some of our assumptions regarding the heat flow estimations might be oversimplified. For example, our investigation, as well as the fold wavelength modeling studies, assumed that the thermal gradient is constant across the lithosphere and neglected the presence of radiogenic heat sources in the crust. However, radiogenic elemental concentrations on the surface of Venus as were measured by the Soviet landers \citep{Surkov1987} are consistent with moderately radiogenic basaltic rocks found on Earth and should contribute for the surface heat flow \cite[see][for a discussion]{Ruiz2019, Karimi2020}. Hence the heat flow near the surface, where the folds formed, should be higher than the global crustal average.  Finally, several studies have also discussed the possibility that the wavelength of tectonic features may not be controlled by the depth of the brittle-ductile transition and could, for example, be associated with intracrustal layering \citep{Montesi2002, Ghent2005, Romeo2011}. Therefore, the observed wavelength might not be purely dependent on the thermal properties of the crust. On the other hand, our simple loading model does not take into account uncorrelated loads and in-plane forces which could could impact the heat flow estimations \cite[e.g.,][]{Mueller1995} given that compressive tectonics probably played an important role in the construction of the crustal plateaus. Nevertheless, studies conducted on orogenic regions on Earth found a good agreement between admittance spectra estimates of the elastic thickness and estimations using more complex forward modeling \citep{McNutt1988}.


\section{Conclusion}\label{sec:conclusion}

We have performed localized admittance modeling at six Venusian crustal plateaus using the spatio-spectral localization technique from \cite{Wieczorek2005,Wieczorek2007}. By testing different compensation scenarios we were able to confirm that most of these features are consistent with being in an Airy isostasy regime. Some extent of flexural support is also accepted given that we found an average upper bound for the elastic thickness of approximately 30 km. The addition of subsurface loads does not have a major influence on the elastic thickness and crustal thickness estimations for most regions.  The average crustal thickness of the plateaus is constrained between about 15 to 34 km, but because of lithospheric deflection the crustal materials can, in several cases, reach down to more than 40 km depth. These values are comparable to the crustal thickness of the continents on Earth. In addition, we were able to estimate that the average crustal thickness of the planet is 22 km when assuming an Airy isostasy, whereas for the surface-loading scenario we found a global average thickness of 17 km. The main discrepancy we found regarding crustal thickness estimations is associated with Phoebe Regio. In this region, when internal loads are not taken into account, we find anomalously high crustal thicknesses that are about 3 times larger than the other plateaus. Only when subsurface loads were added did we find crustal thicknesses that were consistent with the other regions. This indicates that Phoebe is in a different compensation regime compared to other plateaus, being partially supported by a buoyant layer in the mantle.

The elastic thickness and crustal thickness estimations provide some insights on the thermal evolution of the crustal plateaus. Adopting the method introduced by \cite{McNutt1984}, we used the elastic thickness to calculate the heat flow during the period of load emplacement. We then compared these results with other heat flow estimations, such as fold wavelength modeling studies \citep{Brown1997, Resor2021}, and evaluated the possibility of phase transitions and melting at the base of plateaus during their geologic evolution. These analyses indicate that the crustal plateaus formed under higher heat flow conditions compared to the estimate current global average. Melting of deep crustal materials may have happened during the formation of the plateaus, and in some places eclogite may have formed in the deepest crust when the heat flow was lower, potentially leading to crustal delamination and to the recycling of crustal materials into the mantle.

It is clear that a better understanding of the formation of the plateaus will be obtained with new data, including a better gravity model, composition-related measurements  of the surface and higher resolution imaging. For example, if the plateaus are indeed felsic, as thermal emission data possibly indicates \citep{Gilmore2015}, it would reinforce the hypothesis that they are analogous to the continents on Earth. In this context, being able to estimate their bulk crustal density, using gravity techniques applied to the Moon and Mars \cite[e.g.,][]{Wieczorek2012, Broquet2019}, would be extremely valuable. With the currently available gravity data, however, this is not possible. Fortunately, a better gravity model is one of the many datasets that will be obtained by the two planned Venus orbital missions VERITAS \citep{Smrekar2021} and EnVision \citep{Widemann2020} which should have a high enough resolution for constraining the crustal density of plateaus. Undoubtedly, these missions will play fundamental roles towards a better comprehension of the complex geological history of our twin planet.


%
%
%
%

\section*{Data Availability Statement}

A Python routine to compute the admittance flexural model can be found in \cite{maia_2022}. In the same repository the observed localized admittance and correlation are available for each region, as well as the spherical harmonic coefficients for the local crustal thickness maps. In order to perform the localized spectral analysis we used the open source python package pyshtools \citep{Wieczorek2018} and the heat flow estimations were based on the TeHFConversion package from \cite{broquet_2021_4922606}. The spherical harmonic model of the gravity field used in this study can be found on \cite{SJOGRENdataset}, and the topography dataset is from \cite{Wieczorek2015topo}. Both datasets can also be directly loaded in Python using pyshtools. The color maps used in this work are from \cite{crameri2018}.

\section*{Acknowledgments}

We appreciate the helpful discussions with Adrien Broquet. We thank the reviews by Peter James and Sue Smrekar that helped improve this manuscript.

\bibliographystyle{plainnat}
\bibliography{crustalplateaus_ms}

\end{document}